\documentclass[aps,showpacs,nofootinbib,superscriptaddress]{revtex4-1}
\usepackage{epsf}
\usepackage{graphicx}
\usepackage{amsmath}
\usepackage{mathrsfs}

\def\slashchar#1{\setbox0=\hbox{$#1$}
   \dimen0=\wd0 \setbox1=\hbox{/} \dimen1=\wd1
   \ifdim\dimen0>\dimen1 \rlap{\hbox to \dimen0{\hfil/\hfil}} #1
   \else  \rlap{\hbox to \dimen1{\hfil$#1$\hfil}} / \fi}

\newcommand{\bea}{\begin{eqnarray}}
\newcommand{\eea}{\end{eqnarray}}

\begin{document}

\title{Neutrino-induced one-pion production revisited: the $\nu_\mu
  n\to\mu^- n\pi^+$ channel} 

\author{E. Hern\'andez} \affiliation{Departamento de F\'\i sica Fundamental 
e IUFFyM,\\ Universidad de Salamanca, E-37008 Salamanca, Spain} 
\author{J.~Nieves}
\affiliation{Instituto de F\'\i sica Corpuscular (IFIC), Centro Mixto
CSIC-Universidad de Valencia, Institutos de Investigaci\'on de
Paterna, Apartado 22085, E-46071 Valencia, Spain}

\pacs{25.30.Pt,13.15.+g}

\begin{abstract}

Understanding single pion production reactions on free nucleons is the
first step towards a correct description of these processes in nuclei,
which are important for signal and background contributions in current
and near future accelerator neutrino oscillation experiments. In this
work, we reanalyze our previous studies of neutrino-induced one-pion
production on nucleons for outgoing $\pi N$ invariant masses below 1.4
GeV.  Our motivation is to get a better description of the $\nu_\mu
n\to\mu^- n\pi^+$ cross section, for which current theoretical models
give values significantly below data.  This channel is very sensitive
to the crossed $\Delta (1232)$ contribution and thus to spin 1/2
components in the Rarita-Schwinger $\Delta$ propagator.  We show how
these spin 1/2 components are nonpropagating and give rise to contact
interactions. In this context, we point out that the discrepancy with
experiment might be corrected by the addition of appropriate extra
contact terms and argue that this procedure will provide a natural
solution to the $\nu_\mu n\to\mu^- n\pi^+$ puzzle.  To keep our model
simple, in this work we propose to change the strength of the spin 1/2
components in the $\Delta$ propagator and use the $\nu_\mu n\to\mu^-
n\pi^+$ data to constraint its value.  With this modification, we now
find a good reproduction of the $\nu_\mu n\to\mu^- n\pi^+$ cross
section without affecting the good results previously obtained for the
other channels.  We also explore how this change in the $\Delta$
propagator affects our predictions for pion photoproduction and find
also a better agreement with experiment than with the previous model.

\end{abstract}

\maketitle

\section{Introduction}

New and more precise measurements of neutrino cross sections in the
few GeV energy region have renewed interest in a better understanding
of electroweak interactions on nucleons and nuclei. This interest
comes from neutrino oscillation experiments and their need to reduce
systematic errors to achieve the precision goals of the neutrino
oscillation program, making new discoveries, like the CP violation in
the leptonic sector, possible.  Neutrinos are detected through their
interactions with the nuclei that form part of the detectors. For
nuclear physics, this represents a challenge because precise knowledge
of neutrino oscillation parameters requires an accurate understanding
of the detector responses, and it can only be achieved if nuclear
effects are under control~\cite{Gallagher:2011zza,
  Morfin:2012kn,Formaggio:2013kya, Alvarez-Ruso:2014bla,
  Mosel:2016cwa, Katori:2016yel}. Neutrino fluxes used in contemporary
and near future long and short baseline experiments (T2K, NO$\nu$A,
MINER$\nu$A, DUNE, ...) are peaked in the 1--5 GeV energy domain, where
weak pion production becomes one of the main reaction
mechanisms~\cite{Formaggio:2013kya}. Nuclear effects, arising from the
fact that the reaction takes place inside of a nuclear medium, or from
the final-state interactions (FSI) of the produced hadrons through
their path across the nucleus will certainly need to be
incorporated\footnote{Weak pion production in dense matter is strongly
  affected by nuclear corrections, which might not be under
  control. As example of the theoretical difficulties faced, we refer
  the reader to the MiniBooNE flux-folded differential
  $d\sigma/dp_\pi$ cross section data in mineral oil reported in
  Ref.~\cite{AguilarArevalo:2010xt}, which cannot be described by the
  state of the art theoretical calculations of Refs.~\cite
  {Lalakulich:2012cj} and \cite{Hernandez:2013jka}. The latter
  approach is based in the chiral-inspired model of
  Ref.~\cite{Hernandez:2007qq} for weak pion production reaction off
  nucleons, which will be updated in this work. MINER$\nu$A pion production
  data for higher neutrino energies ($E_\nu \sim 4 $ GeV) have recently become
  available~\cite{Eberly:2014mra, Aliaga:2015wva,
    McGivern:2016bwh} and show some appreciable inconsistencies,
  mostly in the magnitude of the cross sections, with MiniBooNE measurements. This
  is an open problem that deserves further discussion. Charged current
  pion production data from T2K will be an important check, since the
  neutrino energy range in this experiment is similar to that of MiniBooNE. }.

Nevertheless, the first requirement to put neutrino induced pion
production on nuclear targets on a firm ground is to have a realistic
model at the nucleon level\footnote{ At this point, we should stress
  that the Rein-Sehgal model~\cite{Rein:1980wg}, used by almost all
  Monte Carlo generators, provides a really poor description of the
  pion electroproduction data on protons~\cite{Graczyk:2007bc,
    Leitner:2008fg}. Indeed, the model underestimates significantly
  the electron data, and it also reveals itself unsatisfactory in the
  axial sector at $q^2 = 0$, where the divergence of the axial current can
  be related to the $\pi N$ amplitude by PCAC (partial conservation of
  the axial current).}. Data on neutrino pion production off nucleons
all come from deuterium bubble chamber experiments carried out in the
1980's at Argonne (ANL)~\cite{Radecky:1981fn} and Brookhaven
(BNL)~\cite{Kitagaki:1986ct} national laboratories. The overall
neutrino-flux normalizations of these measurements have been recently
reanalyzed and corrected in
Refs.~\cite{Wilkinson:2014yfa,Rodrigues:2016xjj}. For antineutrinos,
the measurements are of lower quality, and data on single nucleons are
not available. Most of the models describe the pion production process
by means of the weak excitation of the $\Delta(1232)$ resonance
followed by its strong decay into $N\pi$ ($\Delta-$pole mechanism
depicted in the left-top diagram of Fig.~\ref{fig:npip}), and in some
occasions, incorporate background terms. The major part of the models
includes also the weak excitation of higher resonances as intermediate
states. Vector form factors are fixed from helicity amplitudes
extracted in the analysis of pion electroproduction data, while the
axial couplings are obtained from PCAC~\cite{Alvarez-Ruso:2014bla}.

In this work, we pay a special attention to the $\nu_\mu n\to\mu^-
n\pi^+$ cross section, for which current theoretical models give
values significantly below data. Actually, this channel is certainly
much worse described than the others, $\nu_\mu n\to\mu^- p\pi^0$ and
$\nu_\mu p\to\mu^- p\pi^+$, included in the ANL and BNL data sets.  We
reanalyze our previous study in Ref.~\cite{Hernandez:2007qq} of
neutrino-induced one-pion production on nucleons and show that this
anomaly  could be greatly improved by the addition of appropriate extra
local terms. Such contributions are intimately related to the spin 1/2
degrees of freedom present in the Rarita-Schwinger (RS) $\Delta$
propagator and greatly suppress the crossed $\Delta$ mechanism
(left-bottom diagram in Fig.~\ref{fig:npip}). We find that the use of 
(almost) consistent $\Delta$ couplings~\cite{Pascalutsa:2000kd}, which keep
only the spin 3/2 contribution from the $\Delta$ propagator, leads to
an overall good description of the ANL and BNL data in all three
available charge channels.

The work is organized as follows: After this introduction in
Sec.~\ref{sec:problem},  we briefly review
the most relevant ingredients of model of Ref.~\cite{Hernandez:2007qq}, updated in
Refs.~\cite{Hernandez:2013jka, Alvarez-Ruso:2015eva}, together with
its predictions  for the $\nu_\mu n\to\mu^- n\pi^+$ cross section for
outgoing pion-nucleon invariant masses below 1.4 GeV. Next in Sec.~\ref{sec:rsp}, we describe
the $\Delta(1232)$ propagator used in Ref.~\cite{Hernandez:2007qq},
and show that it is a Green function of the RS equation of motion. We also give
its decomposition into a spin 3/2 part plus the rest. The latter is
a nonpropagating spin 1/2 contribution that gives rise to contact
interactions, at least in the limit of zero $\Delta$ width. In
Sec.~\ref{sec:pascalutsa}, we describe the prescription of 
Ref.~\cite{Pascalutsa:2000kd} to go from inconsistent to consistent
couplings and show the effects of using consistent couplings in the
evaluation of an amplitude where the $\Delta$ appears as an
intermediate state. The extension (modification) of our model is
described in Sec.~\ref{sec:modelext}, and the new results are presented
in Sec.~\ref{sec:results}, where results for pion
photoproduction are also given. The amplitude for this latter process derives from the
vector part of our model for weak pion production, and it
is described in the Appendix.  Finally, in Sec.~\ref{sec:conclusions}
we summarize  the main conclusions of this work.

\section{The model of Refs.~\cite{Hernandez:2013jka, Hernandez:2007qq, 
Alvarez-Ruso:2015eva}: Off diagonal Goldberger-Treiman
  relation, Watson's theorem and the $\nu_\mu n\to\mu^- n\pi^+$ cross
  section}
\label{sec:problem}

In Ref.~\cite{Hernandez:2007qq}, we developed a model for
neutrino-induced one-pion production off the nucleon at low energies
where, besides the dominant $\Delta$ mechanism, we included also
nonresonant contributions required by chiral symmetry. These chiral
background terms were evaluated using a nonlinear SU(2) chiral
Lagrangian and we supplemented them with well-known phenomenological
form factors introduced in a way that respected both CVC (conservation
of the vector current) and PCAC. As for the dominant $\Delta$
contribution, the weak $N\to \Delta$ transition matrix element can be
parametrized in terms of four vector $C^V_{3-6}$ form factors and four
axial $C^A_{3-6}$ ones. $C_6^V$ is exactly zero from CVC, while the
rest of the vector form factors were determined from pion
electroproduction, and for them, we adopted the values in
Ref.~\cite{Lalakulich:2005cs}. Axial form factors are mostly
unknown. In fact, one uses the weak pion production process as a tool
to extract information on the axial nucleon to resonance transition
form factors.  The term proportional to $C_5^A$ is the dominant one.
Assuming the pion pole dominance of the pseudoscalar $C_6^A$ form
factor, PCAC gives its value in terms of $C_5^A$ as
$C^A_6=C^A_5\frac{M^2}{m_\pi^2-q^2}$, where $q^\mu$ is the lepton
transfer four momentum and $M$ ($m_\pi$) the nucleon (pion) mass. We
further adopted Adler's model \cite{Adler:1968tw} in which one has
$C^A_3=0$, $C^A_4=-\frac14 C_5^A$.  We fitted $C_5^A$ to data assuming
a modified dipole parametrization. The experimental data set consisted
of the flux-averaged $q^2$-differential $\nu_\mu p\to \mu^-p\pi^+$
cross section measured at ANL~\cite{Radecky:1981fn}, which
incorporated a kinematical cut $W_{\pi N}<1.4\,$GeV on the invariant
mass of the final nucleon-pion pair.  This was appropriate since our
model ignored the contribution from higher mass resonances. From the
fit we obtained $C_5^A(0)=0.87\pm0.08$. This result was at variance
with the value derived from the off diagonal Goldberger-Treiman
relation (GTR) that predicts $C_5^A(0)\sim 1.15$--$1.20$.

The disagreement with the GTR value got reduced in
Ref.~\cite{Hernandez:2010bx} where, following the work of
Ref.~\cite{Graczyk:2009qm}, we included in our fit total cross
sections measured at BNL~\cite{Kitagaki:1986ct}, and we fully evaluated deuteron effects,
the latter relevant since ANL and BNL data were actually obtained
using a deuterium target.  We had already noticed in
Ref.~\cite{Hernandez:2007qq} that the correct description of BNL cross
sections required larger $C_5^A(0)$ values.  Our preference at the
time for ANL data was due to the fact that they provided absolute
$q^2$-differential cross sections (as opposed to BNL, where only the
shape was given) evaluated with a kinematical cut appropriate for our
model. BNL cross section values are larger and they seemed to be
incompatible with ANL ones. As it has recently been demonstrated in
Ref.~\cite{Wilkinson:2014yfa}, where a reanalysis of both ANL and BNL
data has been conducted, the discrepancies between the two data sets
stem from their respective uncertainties in the neutrino flux
normalization.  In Ref.~\cite{Hernandez:2010bx}, in addition to the ANL
flux-averaged $q^2$-differential $\nu_\mu p\to \mu^-p\pi^+$ cross
section, we included in the fit the three lowest neutrino energy
$\nu_\mu p\to \mu^-p\pi^+$ total cross sections from BNL, and we
considered the uncertainties on the 
 neutrino flux normalizations as fully correlated systematic
errors.  Deuteron effects turned out to reduce the cross section by
some 10\% which agreed with previous estimates in
Refs.~\cite{Graczyk:2009qm,AlvarezRuso:1997jr}. To
compensate this reduction in the cross section, a roughly 5\% larger
$C_5^A(0)$ value was needed. However, it was the consideration in the
fit of BNL cross sections what was responsible for the larger
change in $C_5^A(0)$. Assuming a simpler pure dipole form for $C_5^A$,
we obtained $C_5^A(0)=1.0\pm0.1$, a value closer to the GTR one\footnote{In
  some fits carried out in \cite{Hernandez:2010bx}, we unsuccessfully 
  relaxed 
  Adler's  constraints exploring the possibility of extracting
  some direct information on $C_{3,4}^A(0)$. We showed there that, the
  available low-energy data cannot effectively disentangle the
  different form-factor contributions.}.

In Ref.~\cite{Hernandez:2013jka}, and in order to extend the model to
higher neutrino energies (up to 2 GeV), we added the contribution from
the spin 3/2 nucleon $D_{13}(1520)$ resonance. This is the only resonance, apart from
the $\Delta$, that gives a significant contribution in that energy
region~\cite{Leitner:2008wx}. The corresponding vector and axial form
factors for the $N\to D_{13}$ transition current were taken,
respectively, from fits to results in Refs.~\cite{Leitner:2009zz} and
~\cite{Lalakulich:2006sw}, respectively. A full account of the
$D_{13}(1520)$ contribution can be found in the Appendix
of Ref.~\cite{Hernandez:2013jka}.

\begin{figure}[t!]
\vspace{.5cm}
\makebox[0pt]{\includegraphics[scale=0.6]{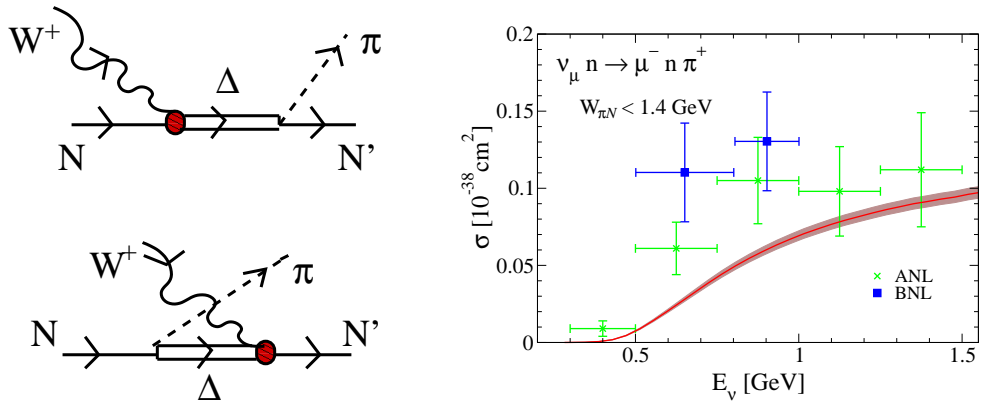}\hspace{1cm}\resizebox{!}{5cm}{\includegraphics{npip_mcfarland.eps}}}
\caption{ Left: Direct
 (top) and crossed (bottom)
  $\Delta(1232)-$pole mechanisms. Right:  $\nu_\mu n\to\mu^-n\pi^+$ total cross section obtained with the
parameters from fit B in Ref.~\cite{Alvarez-Ruso:2015eva} as compared to 
ANL~\cite{Radecky:1981fn} and BNL~\cite{Kitagaki:1986ct} data.
 ANL data and
theoretical results include a cut $W_{\pi N} <1.4\,$GeV in the final pion-nucleon
invariant mass. 
Experimental points include a systematic error due to flux uncertainties (assumed
to be 20\% for ANL and 10\% for BNL data), which
had been added in quadratures to the statistical ones.
Theoretical bands correspond to the variation of the results
 when $C_5^A(0)$ changes within its error interval.}
\label{fig:npip}
\end{figure}

Finally, in Ref.~\cite{Alvarez-Ruso:2015eva}, we partially unitarized
our model by imposing Watson's theorem. This theorem is a result of
unitarity and time-reversal invariance, and it implies that the phase
of the electro or weak pion production amplitude is fully determined
by the strong $\pi N\to\pi N$ interaction elastic phase shifts
$\left[\delta_{L_{2J+1,2T+1}}(W_{\pi N})\right]$. Imposing Watson
restrictions in general is a difficult task, and thus in
\cite{Alvarez-Ruso:2015eva}, we only paid attention to the dominant
spin-3/2 isospin-3/2 positive-parity amplitude, where the direct
excitation of the $\Delta(1232)$ resonance occurs.  Following the
procedure suggested by M.G. Olsson in Ref.~\cite{Olsson:1974sw}, we
introduced independent vector and axial phases (two-dimensional
functions of $q^2$ and $W_{\pi N}$) to correct the interference
between the dominant direct $\Delta$ term  and the nonresonant
background. These extra phases were fixed by requiring that the total
(resonance plus background contributions) amplitude in this dominant
channel had the correct phase $\delta_{P_{33}}(W_{\pi N})$.  Since
this was not possible in a consistent way for all different terms that
contribute to the $P_{33}$ amplitude in the multipolar expansion, we
unitarized only the dominant vector and axial multipoles. Within this
scheme, we performed two different fits in
Ref.~\cite{Alvarez-Ruso:2015eva}. For fit A, we used the same
input data as in Ref.~\cite{Hernandez:2010bx} and described
above. As a consequence of imposing Watson's theorem the interference
between the dominant direct $\Delta$ contribution (left-top diagram of
Fig.~\ref{fig:npip}) and the background
terms changed, and as a result, a larger value ($1.12\pm0.11$) for
$C_5^A(0)$, in agreement now with the GTR prediction, was obtained.  For fit B
we used the results of Ref.~\cite{Wilkinson:2014yfa}.  As already
mentioned, the authors of Ref.~\cite{Wilkinson:2014yfa} reanalyzed ANL
and BNL experiments producing data on the ratio between the
$\sigma(\nu_\mu p\to\mu^- p\pi^+)$ and the charged current
quasielastic (CCQE) cross sections measured in deuterium. In this
way, the flux uncertainties present in the experiments cancel.  They
found a good agreement between the two experiments for these
ratios. Then, by multiplying the cross section ratio by the
theoretical CCQE cross section on the deuteron\footnote{For that
  purpose, they used the prediction from GENIE
  2.9~\cite{Andreopoulos:2009rq}.}, which is well under control, flux
normalization independent pion production cross sections were
extracted. We took advantage of these developments, and for fit B in
Ref.~\cite{Alvarez-Ruso:2015eva}, we considered the new data points.
Since no cut in $W_{\pi N}$ was imposed on this new data, we only used
total cross sections for neutrino energies below 1\,GeV. Besides, to
constrain the $q^2$ dependence of the $C_5^A$ form factor, we also
fitted the shape of the original ANL flux-folded $d\sigma/dq^2$
distribution, where a $W_{\pi N}<1.4\,$GeV cut was
implemented.  For this fit, we obtained $C_5^A(0)=1.14\pm0.07$, similar
to the result from fit A. The quality of the fit, the predictions for
cross sections in other channels, as well as the values of the Olsson
phases needed to satisfy Watson's theorem were very similar in fits A
and B.

The agreement of the theoretical predictions with data was also 
good for the total cross sections in other channels with
one notable exception, the \hbox{$\nu_\mu n\to\mu^-n\pi^+$} reaction shown in 
the right panel of Fig.~\ref{fig:npip}, where theoretical  predictions  lie below 
experimental points. This  is a common problem to other
models~\cite{Sato:2003rq,Kamano:2012id, Nakamura:2015rta, Alam:2015gaa}\footnote{
Note that in Ref.~\cite{Alam:2015gaa}, the theoretical predictions are  below
data when the cut $W_{\pi N} <1.4\,$GeV is implemented. Indeed, this
work uses the  SU(2) chiral model derived in
Ref.~\cite{Hernandez:2007qq}, imposing GTR and  including
smaller contributions from other resonances different to the $\Delta(1232)$ and
the $D_{13}(1520)$.}. A special mention deserves the dynamical model of photo-, electro- and weak
pion production  initially derived in Ref.~\cite{Sato:2003rq}, and that
has been recently  further
refined and extended to incorporate $N^*$ resonances and a larger number of 
meson-baryon states~\cite{Kamano:2012id,Nakamura:2015rta}. Despite its
theoretical and phenomenological robust support, satisfying unitary constrains and 
fulfilling thus Watson's theorem, this model provides  a 
description of  the $\nu_\mu n\to\mu^-n\pi^+$ channel, only slightly better~\cite{sato} to
that shown here in the right panel of Fig.~\ref{fig:npip}. 

As can be deduced from the explicit expressions given in
Ref.~\cite{Hernandez:2007qq}, the $\nu_\mu n\to\mu^-n\pi^+$ reaction
gets a large contribution from the crossed $\Delta$ mechanism
(left-bottom diagram in Fig.~\ref{fig:npip}), and thus it is very
sensitive to the spin 1/2 components present in the RS covariant
$\Delta$ propagator. Indeed, besides the $\Delta$ propagator, the
numerical factors of the (direct $\&$ crossed) $\Delta$ mechanisms are
($\sqrt{3}~ \& ~ 1/\sqrt{3}$ ), ($-\sqrt{2/3}~ \&~ \sqrt{2/3}$), and
($1/\sqrt{3}~ \& ~ \sqrt{3}$) for the $p\pi^+$, $p\pi^0$, and $n\pi^+$
channels, respectively\footnote{Note that the $p\pi^0$ coefficients
  quoted in a similar discussion in Ref.~\cite{Alvarez-Ruso:2015eva}
  were wrong by an overall $-1/\sqrt{2}$ factor.}.  Thus, isospin
invariance implies that the largest (smallest) contribution of the
crossed $\Delta$ mechanism occurs in the $n\pi^+$ ($p\pi^+$) channel,
while the largest (smallest) contribution of the direct $\Delta$
mechanism in contrast is found in the $p\pi^+$ ($n\pi^+$) amplitude.

The RS covariant propagator, with its lower spin components, is
considered to be incorrect in Ref.~\cite{Williams:1985zz}, where the
authors advocate the use of the pure spin 3/2 propagator of Behrends
and Fronsdal~\cite{fronsdal}. The opposite view is adopted in
Ref.~\cite{Benmerrouche:1989uc}, where the pure spin 3/2 propagator is
considered incorrect since it does not satisfy the appropriate Green
function equation. In Ref.~\cite{Aurilia:1969bg}, it is argued that
off shell terms of lower spin can naturally appear in the construction
of propagators, and such terms explain, for instance, the decay of a
spinless pion through an intermediate vector meson, without violating
the conservation law of angular momentum. It is only because the
vector propagator has an off shell spin 0 part that the charged pion
can decay~\cite{Aurilia:1969bg,Benmerrouche:1989uc,Mariano:2012zz}.
What is also true is that those lower spin terms are always
nonpropagating giving rise to pure contact interactions. In
Refs.~\cite{Pascalutsa:1998pw,Pascalutsa:1999zz,Pascalutsa:2000kd}, the
approach is somewhat different. There, the authors arrive at a pure
spin 3/2 contribution from the $\Delta$ propagator by selecting
consistent couplings. These are derivative couplings that preserve the
gauge invariance of the free massless spin 3/2 Lagrangian.  In
Ref.~\cite{Pascalutsa:2000kd}, it is shown how to obtain consistent
couplings from inconsistent ones by just a redefinition of the spin
3/2 field. The difference amount to contact terms that in this
approach are responsible for the contribution of the extra lower-spin
degrees of freedom.  However, as already acknowledged in
Ref.~\cite{Pascalutsa:1998pw}, and very recently reanalyzed in
Ref.~\cite{Badagnani:2015rwj}, consistent couplings cannot be kept in
the presence of electromagnetic interactions. This is so since any
derivative on the $\Delta$ field gives rise through minimal
substitution to a new nonderivative term.

Our approach to this
problem is conceptually different, based on the perspective of an  effective
field theory, and it is motivated by the discussion in
Ref.~\cite{Pascalutsa:2000kd}. In this latter reference, it is argued
that 
i) the use of consistent or inconsistent couplings will provide the
same physical predictions as far as all relevant contact terms allowed
by the underlying symmetries are included in both cases, and ii) the
strength of the contact terms will have to be fitted to
experiment. According to this, in this work, we propose a minimal
modification of our model, in which the contact terms that derive from
the spin 1/2 part of the $\Delta$ propagator are multiplied by an
extra parameter (low energy constant), that will be fitted to 
data.

\section{Rarita-Schwinger propagator}
\label{sec:rsp}

The RS  Lagrangian of the free massive spin 3/2
 reads~\cite{Pascalutsa:2000kd} [we particularize for the
$\Delta(1232)$ resonance case],
\begin{equation}
{\cal L}_{\rm RS} = \bar\Psi_\mu \Lambda^{\mu\nu}\Psi_\nu, \quad
\Lambda^{\mu\nu} = \left(\gamma^{\mu\nu\alpha}i \partial_\alpha - M_\Delta
\gamma^{\mu\nu}\right) = \frac12 \left \{(i
\slashchar{\partial}-M_\Delta),\gamma^{\mu\nu}\right\}_+ 
\label{eq:lagran}
\end{equation} 
where $\Psi_\mu$ represents the RS field for the $\Delta$ and
\bea
&&\gamma_{\beta\nu\alpha}=\frac{1}2\{\gamma_{\beta\nu},\gamma_\alpha\}_+=-i
\epsilon_{\beta\nu\alpha\rho}\gamma^\rho\gamma_5
,\ \ \gamma_{\beta\nu}=\frac12[\gamma_\beta,\gamma_\nu].
\eea
with $\epsilon_{0123}=+1$ and $g^{\mu\nu}=(1,-1,-1,-1)$. The
Lagrangian in Eq.~\eqref{eq:lagran} corresponds to the parameter
$A=-1$ in the discussion of Eq.~(2) of Ref.~\cite{Benmerrouche:1989uc}
(note that the physical properties of the free field are independent of
this parameter).  The Euler-Lagrange equation reads
\bea
\Lambda^{\mu\nu}\Psi_\nu& =& (\gamma^{\mu\nu\alpha}i \partial_\alpha - M_\Delta
\gamma^{\mu\nu})\Psi_\nu =
-\left[(i\slashchar{\partial}-M_\Delta)g^{\mu\nu}+ \gamma^\mu(i\slashchar{\partial}+M_\Delta)\gamma^\nu-i(\gamma^\mu\partial^\nu+\partial^\mu\gamma^\nu)\right]\Psi_\nu=0
\eea
which leads to the set of equations
\begin{equation}
(i\slashchar{\partial}-M_\Delta)\Psi_\nu=0, \quad
  \partial^\nu\Psi_\nu=0, \quad \gamma^\nu\Psi_\nu=0
\end{equation}
The corresponding RS propagator is
\begin{equation}
G_{\mu\nu}(p_\Delta)=
\frac{P_{\mu\nu}(p_\Delta)}{p_\Delta^2-M_\Delta^2+ i M_\Delta
  \Gamma_\Delta}  \label{eq:delta-prop} 
\end{equation}
\begin{equation}
P^{\mu\nu}(p_\Delta)= - (\slashchar{p}_\Delta + M_\Delta) \left [ g^{\mu\nu}-
  \frac13 \gamma^\mu\gamma^\nu-\frac23\frac{p_\Delta^\mu
  p_\Delta^\nu}{M_\Delta^2}+ \frac13\frac{p_\Delta^\mu
  \gamma^\nu-p_\Delta^\nu \gamma^\mu }{M_\Delta}\right]
\end{equation}
In the zero width limit ($\Gamma_\Delta=0$, i.e., when dealing with a
stable particle), the above propagator gives the Green function of the
RS equation of motion
\bea
\Lambda_{\alpha\beta}G^\beta_\delta(x) =
g_{\alpha\delta}\delta^4(x),
\eea
 with $G^{\mu\nu}(x)$ the Fourier's
transform of $G^{\mu\nu}(p_\Delta)$. This result follows trivially  from
\bea
(\gamma_{\mu\nu\alpha}p_{\Delta}^\alpha - M_\Delta
\gamma_{\mu\nu})P^{\nu\beta}(p_\Delta)=
(p^2_\Delta-M^2_\Delta)g_{\mu}^\beta,
\eea
 which can be obtained after a
little of Dirac algebra.

The $P^{\mu\nu}$ operator can be rewritten as~\cite{Pascalutsa:1999zz}
\begin{eqnarray}
P_{\mu\nu}(p)&=& P_{\mu\nu}^\frac32(p)+
(p^2-M_\Delta^2)\left[\frac2{3M_\Delta^2}(\slashchar{p} +
M_\Delta)\frac{p_\mu
  p_\nu}{p^2 }-\frac{1}{3 M_\Delta}\left(\frac{p^\rho
  p_\nu\gamma_{\mu\rho}}{p^2}+ \frac{p^\rho
  p_\mu\gamma_{\rho\nu}}{p^2}\right)\right], \label{eq:p-p32}
  \end{eqnarray}
  with
\begin{eqnarray}
P_{\mu\nu}^\frac32(p)&=& - (\slashchar{p}+ M_\Delta) \left [ g_{\mu\nu}-
  \frac13 \gamma_\mu\gamma_\nu -
  \frac{1}{3p^2}\left(\slashchar{p}\gamma_\mu p_\nu +
  p_\mu\gamma_\nu\slashchar{p}\right)\right].
  \end{eqnarray}
$P_{\mu\nu}^\frac32(p)$  satisfies the relations
\begin{eqnarray}
0=[\slashchar{p},P_{\mu\nu}^\frac32(p)] =  p^\mu P_{\mu\nu}^\frac32(p)
= P_{\mu\nu}^\frac32(p)p^\nu =  \gamma^\mu P_{\mu\nu}^\frac32(p) = P_{\mu\nu}^\frac32(p)\gamma^\nu, \quad
P_{\mu\nu}^\frac32(p)[P^\frac32(p)]^{\nu\rho}= -(\slashchar{p}+ M_\Delta) [P^\frac32(p)]_\mu^{\rho}
\end{eqnarray}
from where one concludes that $P_{\mu\nu}^\frac32$ is the spin-3/2
projection operator. 

Finally, we would like to stress that Eq.~\eqref{eq:p-p32} shows that
in the RS propagator of Eq.~\eqref{eq:delta-prop}, only   the spin-3/2 degrees of freedom propagate, while
the controversial spin-1/2 contributions give rise to contact
background terms. (This is strictly true in the zero  width limit
where the factor $(p^2-M_\Delta^2)$ in Eq.~\eqref{eq:p-p32} cancels
the denominator of the $\Delta$ propagator.) As we will discuss below,
the total strength of the contact terms is undetermined in an effective
chiral expansion, and it needs to be determined from experiment.
\section{Consistent $\Delta$ interactions: the prescription of Ref.~\cite{Pascalutsa:2000kd}}
\label{sec:pascalutsa}
The kinetic term of the free RS Lagrangian in Eq.~(\ref{eq:lagran}) is invariant
under the gauge transformation
\begin{equation}
\Psi_\mu(x) \to \Psi_\mu(x)+ \partial_\mu \epsilon (x), \label{eq:gauge}
\end{equation}
with $\epsilon (x)$ a spinor. It is argued in Refs.~\cite{Pascalutsa:1998pw,
Pascalutsa:1999zz} that any interaction term  that respects this symmetry 
does not change the degrees of freedom content of the free theory, where the
constraints on  $\Psi_\mu(x)$ guarantee that it indeed describes spin 3/2
particles. Couplings respecting this symmetry 
are called consistent ones.
In the case of linear couplings of the form 
\bea 
{\cal L}_{\rm int}= g\, \bar \Psi_\beta J^\beta +h.c., 
\eea 
where $J^\mu$ is any current coupled to
$\Delta$, the invariance of the Lagrangian under the gauge transformation
requires the current $J^\mu$ to be conserved. 
If that is not the case, the
coupling is called inconsistent.
The transformation of this latter coupling into a consistent one can be achieved
via a redefinition of the $\Delta$ field
\bea
\Psi_\mu \to \Psi_\mu + g\, \xi_\mu. 
\eea
This transformation modifies the linear coupling
\bea
{\cal L}'_{\rm int}=g\, \bar \Psi_\beta (J^\beta +\Lambda^{\beta\nu}\xi_\nu)+h.c.
\label{eq:nlc}
\eea
and gives rise to an additional contact interaction Lagrangian, ${\cal L}_C$,
independent of the RS field (see Ref.~\cite{Pascalutsa:2000kd} for details on
${\cal L}_C$). By selecting
\begin{equation}
\xi_\mu=\left(M_\Delta \gamma^{\mu\nu}\right)^{-1}J^\nu=
-\frac{1}{M_\Delta} {\cal   O}^{(-1/3)}_{\mu\nu}J^\nu,
\end{equation}
where
\bea
 &&{\cal
O}^{(x)}_{\nu\mu}=g_{\nu\mu}+x\gamma_\nu\gamma_\mu,
\eea
one has that the new total current coupled to the $\Delta$ is
\begin{equation}
{\cal J}^\beta = J^\beta + \Lambda^{\beta\nu}\xi_\nu= \gamma^{\beta\nu\alpha}i\partial_\alpha\xi_\nu =  
-\frac{i}{M_\Delta} 
\gamma^{\beta\nu\alpha} {\cal   O}^{(-1/3)}_{\nu\rho}\partial_\alpha
J^\rho \label{eq:newJ}, 
\end{equation}
which is indeed conserved.

Apart from a total
divergence of no consequence, Eq.~(\ref{eq:nlc}) can be rewritten as
\bea
{\cal L}'_{\rm int}= i\frac{g}{M_\Delta}\partial_\alpha \bar\Psi_\beta\,\gamma^{\alpha\beta\nu}\,{\cal
O}^{(-1/3)}_{\nu\rho}\,J^\rho+h.c.
\eea
This is 
 the prescription described in Ref.~\cite{Pascalutsa:2000kd} to transform an
 inconsistent coupling into a consistent one. 
The description in terms of the original ${\cal L}_{\rm int}$ or the modified ${\cal L}'_{\rm int}+{\cal
  L}_C $ Lagrangians is equivalent at the level of the $S$ matrix. 
  
It is further argued in Ref.~\cite{Pascalutsa:2000kd} that, within
chiral perturbation theory (ChPT), any linear spin-3/2 coupling is
acceptable. This is so since the additional ${\cal L}_C$ contact
terms, which provide the equivalence between inconsistent and
consistent couplings, have to be included in both situations with
arbitrary coefficients that have to be fitted to some experimental input. Thus, it is
only the value of the coefficients of the contact terms that change.
In this respect, the spin-1/2 contributions in the RS propagator, that
give rise to pure contact terms, can be totally eliminated and their
effects reabsorbed into the values of some of the low-energy constants
of the additional zero-range couplings.  According to
Ref.~\cite{Pascalutsa:2000kd}, it is preferable to use  consistent
interaction terms, supplemented with the adequate contact
interactions,  in the analysis of the separate contributions due to
spin 3/2 degrees of freedom versus the rest.

\begin{figure}[h!]
\vspace{.5cm}
\begin{center}
\resizebox{!}{2.5cm}{\includegraphics{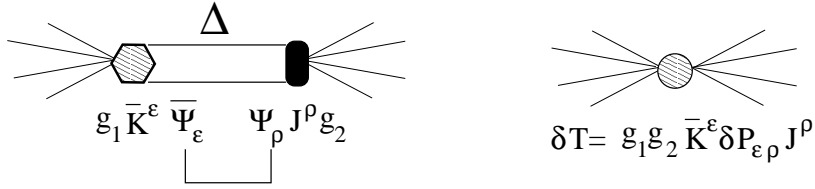}}
\end{center}
\caption{ Left: Reaction mechanism where a $\Delta$ is excited and later on
  it  decays into some final
particles. Right: Contact term that accounts for the difference when
the diagram depicted in the left panel is evaluated using 
  consistent or  inconsistent $\Delta$ couplings.}
\label{fig:consistent}
\end{figure}

To see the effect of the use of consistent interactions, let us
consider a process driven by the excitation of the $\Delta$ and its
subsequent decay into some final particles. This mechanism is depicted
in the left panel of Fig.~\ref{fig:consistent}, and it is determined by
the currents $\bar K^\epsilon$ and $J^\rho$ that couple the $\Delta$
to the initial and final particles, respectively, and that we assume
to be of the inconsistent type.
%
%
In the zero width limit, the amplitude for the process would be 
\begin{equation}
T= g_1g_2 \bar K^\epsilon \frac{P_{\epsilon\rho}}{p_\Delta^2-M^2_\Delta}J^\rho,\label{eq:t}
\end{equation}
while using the consistent currents $\bar {\cal K}^\epsilon$ and
${\cal J}^\rho$, one would get
\begin{eqnarray}
T_{ consistent} &= &g_1g_2\, \bar {\cal K}^\epsilon
\frac{P_{\epsilon\rho}}{p_\Delta^2-M^2_\Delta}{\cal J}^\rho\\
 &= &g_1g_2\,
\frac{p_{\Delta\eta}p_{\Delta\sigma}}{M^2_\Delta}\bar K^\epsilon 
{\cal
  O}^{(-1/3)}_{\epsilon\mu}\gamma^{\mu\eta\alpha}\frac{P_{\alpha\beta}}{p_\Delta^2-M^2_\Delta}\gamma^{\beta\sigma\nu}{\cal
  O}^{(-1/3)}_{\nu\rho}J^\rho \\
&=&g_1g_2 \bar K^\epsilon \frac{p^2_\Delta}{M^2_\Delta}
\frac{P_{\epsilon\rho}^{\frac32}}{p_\Delta^2-M^2_\Delta}
J^\rho.\label{eq:tcons}
\end{eqnarray}
This result follows from the antisymmetry of the $\gamma^{\mu\eta\alpha}$
tensor that guaranties that,
\begin{eqnarray}
p_{\Delta\eta}p_{\Delta\sigma} {\cal
  O}^{(-1/3)}_{\epsilon\mu}\gamma^{\mu\eta\alpha}P_{\alpha\beta}\gamma^{\beta\sigma\nu}{\cal
  O}^{(-1/3)}_{\nu\rho} &=& p_{\Delta\eta}p_{\Delta\sigma} {\cal
  O}^{(-1/3)}_{\epsilon\mu}\gamma^{\mu\eta\alpha}P_{\alpha\beta}^{\frac32}\gamma^{\beta\sigma\nu}{\cal
  O}^{(-1/3)}_{\nu\rho}\nonumber \\ 
&=& -  p_{\Delta\eta}p_{\Delta\sigma} {\cal
  O}^{(-1/3)}_{\epsilon\mu}\gamma^{\mu\eta\alpha}(\slashchar{p}_\Delta+M_\Delta){\cal
  O}^{(-1/3)}_{\alpha\beta}
\gamma^{\beta\sigma\nu}{\cal
  O}^{(-1/3)}_{\nu\rho}\label{eq:mesh},
\end{eqnarray}
and some further Dirac algebra\footnote{In Eq.~\eqref{eq:mesh}, the
  $g_{\epsilon \mu}$ tensor in ${\cal
  O}^{(-1/3)}_{\epsilon\mu}$ gives the final result, $ p^2_\Delta
  P_{\epsilon\rho}^{\frac32}$,  while the
  $\gamma_\epsilon \gamma_\mu$ part produces an antisymmetric tensor in
  the $\eta$ and $\sigma$ indices whose contribution vanishes
  when contracted with the symmetric
  $p_{\Delta\eta}p_{\Delta\sigma}$ term.}.
 
By comparing Eqs.~\eqref{eq:t} and \eqref{eq:tcons}, we see that  the use 
of  consistent couplings induces  the replacement
\begin{equation}
P_{\epsilon\rho} \leftrightarrow \frac{p^2_\Delta}{M_\Delta^2}P_{\epsilon\rho}^{\frac32}
\end{equation}
in the Feynman amplitudes. Note that the factor $p^2_\Delta$ in front
of $P_{\epsilon\rho}^{\frac32}$ corrects for the ill-defined infrared
behaviour of the latter operator.  From Eq.~\eqref{eq:p-p32} we see 
that $P_{\epsilon\rho}$ and
$P_{\epsilon\rho}^{\frac32}$ differ in terms that vanish on shell
($p^2_\Delta=M^2_\Delta$), 
\bea
P_{\epsilon\rho} -
\frac{p^2_\Delta}{M_\Delta^2}P_{\epsilon\rho}^{\frac32} & = &
(p^2_\Delta- M_\Delta^2) \delta P_{\epsilon\rho}(p_\Delta) \\
\delta P_{\epsilon\rho}(p_\Delta)& = &
\frac1{M_\Delta^2}(\slashchar{p}_\Delta +
M_\Delta)\left(g_{\epsilon\rho}-\frac13\gamma_\epsilon\gamma_\rho\right)+
\frac1{3M_\Delta^2}\left(p_{\Delta\,\epsilon}\gamma_\rho-
p_{\Delta\,\rho}\gamma_\epsilon\right)
\label{eq:def-deltaP}
\eea
and thus, the amplitudes $T$ and $T_{ consistent}$
differ in a contact (nonpropagating) term $\delta T$,
\bea
T &=& T_{ consistent} + \delta T, \qquad \delta T =g_1g_2 \bar K^\epsilon \delta P_{\epsilon\rho}J^\rho
\eea
The discussion above amounts to admit that the
actual size of a contact term like $\delta T$ is in fact undetermined, since 
the contact terms that appear in the effective chiral
expansion are not fixed, and need to be fitted to experiment. Hence the use of
consistent or  inconsistent $\Delta$ couplings should not
produce any difference, as long as the needed contact terms
are phenomenologically determined.

\subsubsection{The $\pi N\Delta$ coupling}
\label{sec:piNDelta}
For the case of the $\pi N\Delta$ coupling, in Ref.~\cite{Hernandez:2007qq} we took 
\bea
{\cal L}_{\pi N \Delta}=\frac{f^*}{m_\pi}\bar\Psi_\beta\, \vec T^\dagger\, \Psi\,\partial^\beta\vec\phi+h.c.
\label{eq:pindel_incon}
\eea 
with $f^*$ the strong coupling constant, $m_\pi$ the
pion mass, $\Psi$ and $\vec \phi$ the nucleon and pion
fields\footnote{In our convention, $\phi=(\phi_x-{\rm i}\,
  \phi_y)/\sqrt{2}$ creates a $\pi^-$ from the vacuum or annihilates a
  $\pi^+$, whereas the $\phi_z$ field creates or annihilates a
  $\pi^0$.}, and $\vec T^\dagger$ the isospin $1/2\to 3/2$ transition
operator defined such that its Wigner-Eckart reduced matrix element is
equal to one. The $\Delta$ width that results from the above vertex,
assuming an onshell $\Delta$ at rest and with mass $W_{\pi N}$, i.e.,
 $p_\Delta^\mu=(W_{\pi N},\vec{0}\,)$, is given
by\footnote{In the expression of Eq.~(45) of
  Ref.~\cite{Hernandez:2007qq}, the factor $(E+M)/2W_{\pi N}$ was
  approximated by $M/W_{\pi N}$.},
\bea 
\Gamma_{\Delta\to
  N\pi}(W_{\pi
  N})=\frac1{6\pi}\Big(\frac{f^*}{m_\pi}\Big)^2\frac{E+M}{2W_{\pi
    N}}\,k_\pi^3\,\Theta\left(W_{\pi
    N}-M-m_\pi \right)
\label{eq.deltawidth-old}
\eea 
where $M,E$ and $k_\pi$ are the mass and energy of the final nucleon
and the final pion momentum, respectively in the $\Delta$ rest frame.  Using isospin
averaged masses and the value $\Gamma_{\Delta\to
  N\pi}(M_\Delta)=117\,$MeV~\cite{pdg} we obtain $f^*=2.15$ to be
compared to the value 2.14 that we have been using so far.
The use of a  consistent coupling would lead to the inclusion of an
additional multiplicative factor $W_{\pi N}^2/M^2_\Delta$.
%
%

To end this section, we would like to devote a few words  to the
use of a more general $\pi N\Delta$ interaction of the form~\cite{Benmerrouche:1989uc}
\bea
\frac{f^*}{m_\pi}\bar\Psi_\beta\, \vec T^\dagger\,
(g^{\beta\alpha}+z\gamma^\beta\gamma^\alpha)\Psi\,\partial_\alpha\vec\phi+h.c.\label{eq:defz}
\eea
In diagrams with an intermediate $\Delta$, and because
$P^{\frac32}_{\alpha\beta}\gamma^\beta$ = 0, the $z$ term will always
give rise to contact contributions which, as argued above, 
need to be phenomenologically determined. Hence, without
lost of generality, one can ignore these off shell terms as far as all relevant contact
interactions are taken into account\footnote{In this context, the inconsistency
 between the free $\Delta$ propagator and the $\pi N\Delta$ Lagrangian referred
  to in Refs.~\cite{Mariano:2012zz, Barbero:2013eqa}
 would no longer be relevant.}.

\section{Extension of the model of Refs.~\cite{Hernandez:2007qq,
 Hernandez:2013jka, Alvarez-Ruso:2015eva} }
\label{sec:modelext}

Aiming at improving the description of the $\nu_\mu n\to\mu^- n\pi^+$
channel, we open the possibility of supplementing the model of
Refs.~\cite{Hernandez:2007qq, Hernandez:2013jka,Alvarez-Ruso:2015eva} with some
additional contact terms. To keep the model simple, we introduce just
one undetermined  low energy constant (LEC), $c$, that enters in a 
modification of the $\Delta$ propagator compatible with ChPT, 
\begin{eqnarray}
\frac{P_{\mu\nu}(p_\Delta)}{p_\Delta^2-M_\Delta^2} \to
\frac{P_{\mu\nu}(p_\Delta)+ c \left(P_{\mu\nu}(p_\Delta) -
\frac{p^2_\Delta}{M_\Delta^2}P_{\mu\nu}^{\frac32}(p_\Delta)\right)}{p_\Delta^2-M_\Delta^2}
=  \frac{P_{\mu\nu}(p_\Delta)}{p_\Delta^2-M_\Delta^2} + c\, \delta P_{\mu\nu}(p_\Delta) \label{eq:def-LEC}
\label{eq:modifipropa}
\end{eqnarray} 
with the operator $\delta P_{\mu\nu}$ defined in
Eq.~\eqref{eq:def-deltaP}.  The introduction of this LEC induces two
new terms in the model that come from the direct ($\Delta\rm P$) and
crossed $\Delta$ pole ($\rm C\Delta\rm P$) amplitudes. [Note that
there no exists an unequivocal relation between the LEC $c$ and the
parameter $z$ introduced in Eq.~(\ref{eq:defz}), and thus effects
produced by the latter cannot be completely accounted by the inclusion
of these two new terms.]

So far, the values $c= 0$ and $c=-1$ would correspond to
the use of  inconsistent and  consistent
$\Delta$ couplings. We now reintroduce in the denominator of 
the propagator in  Eq.~\eqref{eq:def-LEC} the imaginary 
part $ i M_\Delta \Gamma_\Delta$,
where for $\Gamma_\Delta$ we  use Eq.(\ref{eq.deltawidth-old}) with the
new $f^*$ value. Note that the width is 
zero for the $\rm C\Delta\rm
P$ term, while we expect the direct $\Delta\rm P$ contribution to be
largely dominated by the resonant propagator, being there the
influence of the $\delta P_{\mu\nu}$ term quite small. However, we
foresee that the contribution of this latter term could be relevant in 
the $\rm C\Delta\rm  P$ amplitude, because in that case the $\Delta$
is largely off shell. 

It is worth stressing that the nondiagonal GTR is not
affected by the changes and it predicts
\bea
C_5^A(0)=\sqrt{\frac23}\,\frac{f_\pi}{m_\pi}\,f^*,
\eea
that for $f_\pi=93.2\,$MeV and the isospin averaged $m_\pi$ value that we use
results in $C_5^A(0)=1.19$.\\

In principle one could also modify the $D_{13}(1520)$ terms included in
our model (see Ref.~\cite{Hernandez:2013jka}) along the lines
described above and introduce an extra parameter. However, since the 
$D_{13}(1520)$ exchange contributions play a minor
role, the effect of these latter modifications would
be much less important, and we shall ignore them.\\

With the modification  in the $\Delta$ contributions, we  repeat 
the fit B carried out in Ref.~\cite{Alvarez-Ruso:2015eva}. There is a
total of four best fit parameters: the LEC $c$, $C_5^A(0)$, and
$M_{A\Delta}$, that determine  the $C_5^A(q^2)$ axial form factor for
which we assume a dipole form
\bea C_5^A(q^2)=\frac{C_5^A(0)}{\left(1-q^2/M_{A\Delta}^2\right)^2},
\eea and the normalization parameter $\beta$ of the $\nu_\mu
p\to\mu^-p\pi^+$ ANL differential cross section introduced in
Ref.~\cite{Alvarez-Ruso:2015eva}. In addition and to increase the sensitivity on the
new $c$ parameter, we now also include in the fit data for the
$\nu_\mu n\to\mu^-n\pi^+$ reaction. We thus minimize the following
$\chi^2$
\bea
\chi^2&=& \left\{\sum_{i\in {\rm ANL}}\left ( \frac{\beta d\sigma/dQ_i^2|_{\rm
    exp}-d\sigma/dQ_i^2|_{\rm th}}{\beta \Delta(d\sigma/dQ_i^2|_{\rm
    exp}) }\right)^2 + \sum_{i\in {\rm ANL}}\left ( \frac{\sigma_i|_{\rm
    exp}-\sigma_i|_{\rm th}}{ \Delta(\sigma_i|_{\rm
    exp}) }\right)^2+  \sum_{i\in {\rm BNL}}\left ( \frac{\sigma_i|_{\rm
    exp}-\sigma_i|_{\rm th}}{ \Delta(\sigma_i|_{\rm
    exp}) }\right)^2\right\}_{\nu_\mu p\to\mu^- p\pi^+}\nonumber\\
&&+\left\{\sum_{i\in {\rm ANL}}\left ( \frac{\sigma_i|_{\rm
    exp}-\sigma_i|_{\rm th}}{ \Delta(\sigma_i|_{\rm
    exp}) }\right)^2\right\}_{\nu_\mu n\to\mu^- n\pi^+}
    \label{eq:newchi} \,,
\eea
where $Q^2=-q^2$. The $d\sigma/dQ^2$ differential cross section values
are the flux averaged measurements carried out at
~\cite{Radecky:1981fn} (ANL) and they contain a $W_{\pi N}<1.4\,$GeV
cut in the final pion-nucleon invariant mass. This data set serves the
purpose of constraining the $q^2$ dependence of the $C_5^A(q^2)$ axial
form factor.
The role played by the parameter $\beta$ is to allow fitting only the shape of this 
distribution. The total $\nu_\mu p\to\mu^- p\pi^+$
 ANL and BNL  cross sections included in the fit are collected in Table II of 
 Ref.~\cite{Alvarez-Ruso:2015eva}. They have been taken from the reanalysis of
 Ref.~\cite{Wilkinson:2014yfa}, where flux uncertainties in the original 
 ANL and BNL data have been eliminated. Since they do not include a cut in 
 $W_{\pi N}$, we  only consider cross sections for neutrino energies 
 $E_\nu \le 1$ GeV. Finally, for the total $\nu_\mu n\to\mu^- n\pi^+$ cross
 section, we take also the
 results of the reanalysis of the ANL data conducted in Ref.~\cite{Rodrigues:2016xjj}
 and shown in Table~\ref{tab:npip}. In this latter case, the data do contain a 
 $W_{\pi N}<1.4\,$GeV cut.
 \begin{table*}
\caption{$\nu_\mu n\to\mu^- n\pi^+$ ANL cross sections (in units of $10^{-38}$ 
cm$^2$) taken from the
  reanalysis of Ref.~\cite{Rodrigues:2016xjj}. A  $W_{\pi N}<1.4\,$GeV cut has
  been applied to obtain the data.}
\vspace{.3cm}
\begin{tabular}{c|cc|c}
\hline\hline
 $E_\nu$ (GeV) & $\sigma|_{\rm
    exp}$ & $\Delta(\sigma|_{\rm
    exp})$ & Expt. \\\hline  
0.400&    0.010&  0.006& ANL\\
0.625&    0.070&  0.014& ANL\\
0.875&    0.121&  0.022& ANL\\
1.125&    0.110&  0.024& ANL\\
1.375&    0.122&  0.033& ANL\\
\hline\hline
\end{tabular}
\label{tab:npip}
\end{table*}
As in Ref.~\cite{Alvarez-Ruso:2015eva}, we consider deuterium
 effects and Adler's constraints ($C_3^A=0,\;C_4^A=-C_5^A/4$)
  on the axial form factors. Besides, Olsson's approximate
 implementation of Watson's theorem, as described in 
 Ref.~\cite{Alvarez-Ruso:2015eva}, is also taken into account.

\section{Results}
\label{sec:results}
\subsection{Pion production by neutrinos}
The best fit parameters resulting from the new fit are
\bea
C_5^A(0)&=&1.18\pm0.07,\ M_{A\Delta}=950\pm 60\,{\rm
  MeV},\ c=-1.11\pm0.21,\ \beta=1.23\pm0.08.\label{eq:fit-results}
\eea
 The new $\chi^2/dof=1.1$ is
dominated by the $\nu_\mu n\to\mu^-n\pi^+$ reaction that gives rise to
about  $75\%$ of the
total.  $C_5^A(0)$ is now larger by 3.5\% than that found in 
Ref.~\cite{Alvarez-Ruso:2015eva}, and it is in excellent agreement with the GTR value. 
The $\beta$ parameter is a measure of the neutrino flux uncertainty 
in the ANL experiment. Its value is in agreement with the 20\% uncertainty
 assumed for our fit A in  Ref.~\cite{Alvarez-Ruso:2015eva} and the fits 
 in Refs.~\cite{Graczyk:2009qm,Hernandez:2010bx}. 

\begin{figure}
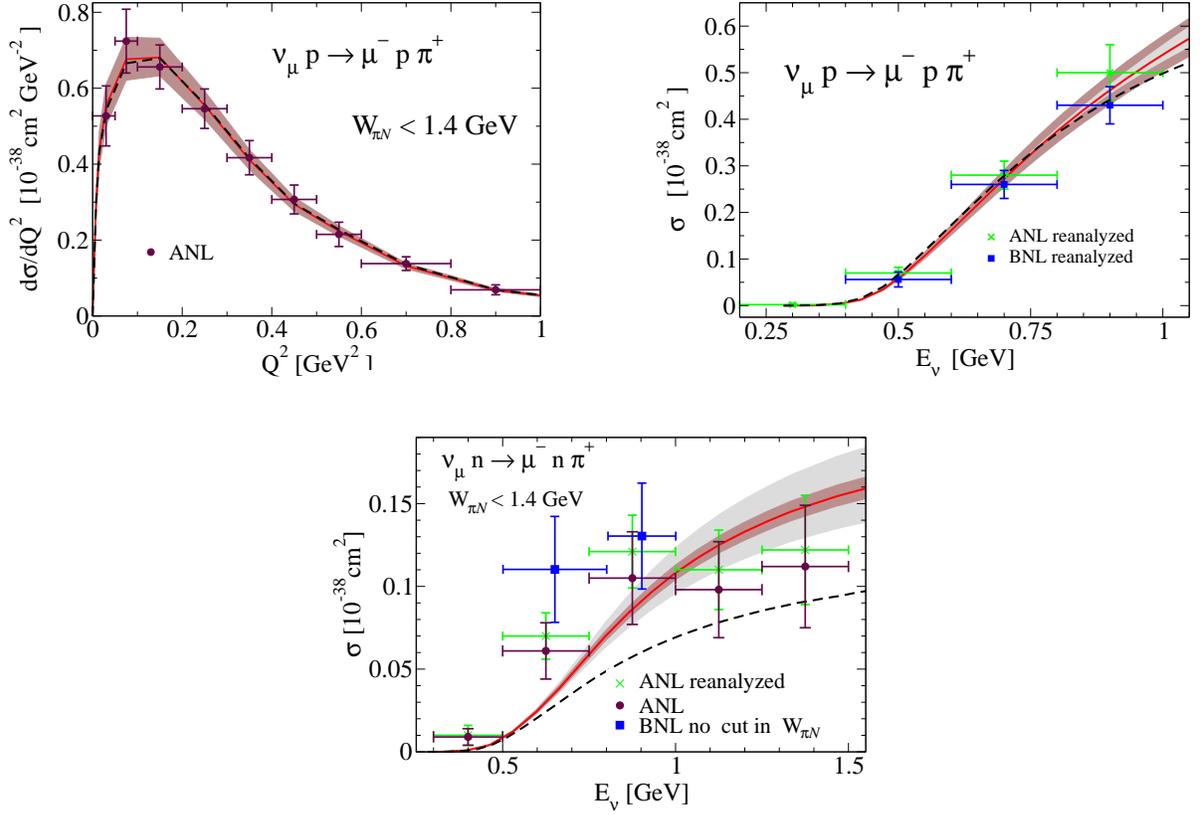

\vspace{.35cm}
\resizebox{!}{5cm}{\includegraphics{dsigdq2_juan2.eps}}\hspace*{1.35cm}
\resizebox{!}{5cm}{\includegraphics{sigma_juan2.eps}}\vspace*{.75cm}
\resizebox{!}{5cm}{\includegraphics{npip_juan2.eps}}
\caption{Theoretical results for the shape of the flux-folded
  differential $d\sigma/dQ^2$ (upper left panel) and total $\nu_\mu
  p\to \mu^- p\pi^+$ (upper right panel) and $\nu_\mu n\to\mu^-n\pi^+$
  (bottom panel) cross sections compared to data from
  ANL~\cite{Radecky:1981fn} (upper left panel) and the reanalyses
  of  Refs.~\cite{Wilkinson:2014yfa} (upper right panel) and
  \cite{Rodrigues:2016xjj} (bottom panel). In the bottom panel, we
  also show the original ANL~\cite{Radecky:1981fn}  and
  BNL~\cite{Kitagaki:1986ct} 
  data.  Red solid and black dashed lines
  show  the results obtained
  in this work, obtained using the best fit parameters
  of Eq.~\eqref{eq:fit-results},  and those derived from fit B of
  Ref.~\cite{Alvarez-Ruso:2015eva}, respectively. In the upper left and bottom
  panels, ANL data (both original and reanalyzed) and
  theoretical results include a  $W_{\pi N}<1.4\,$GeV cut  in the final
  pion-nucleon invariant mass.  Brown (gray) theoretical bands  
  account for the variation of the results
  when $C_5^A(0)$ (LEC $c$) changes within its error interval
  given in Eq.~\eqref{eq:fit-results}. ANL reanalyzed
  cross sections have no systematic errors due to flux uncertainties. Besides,
  theoretical results in the upper left panel have been divided by
  $\beta=1.23$, accounting for flux uncertainties [see
    Eq.~\eqref{eq:newchi}]. Deuteron effects have been taken into
  account as explained in Ref.~\cite{Hernandez:2010bx}. }
\label{fig:nuevofit}
\end{figure}

In Fig.~\ref{fig:nuevofit}, we compare  the fitted data and the 
new theoretical results.  For  comparison we also show the results
from fit B carried out in   
 Ref.~\cite{Alvarez-Ruso:2015eva}. The shape for the flux averaged differential 
  cross section $d\sigma/dQ^2$ 
  $\nu_\mu p\to \mu^- p\pi^+$ is shown in the upper left panel. Both 
  fits give
    almost identical results, as it is also the case for the total 
   $\nu_\mu p\to \mu^- p\pi^+$ cross section, depicted in the upper
    right panel, where some
  minor differences can be only seen for the larger neutrino energies. 
\begin{figure}
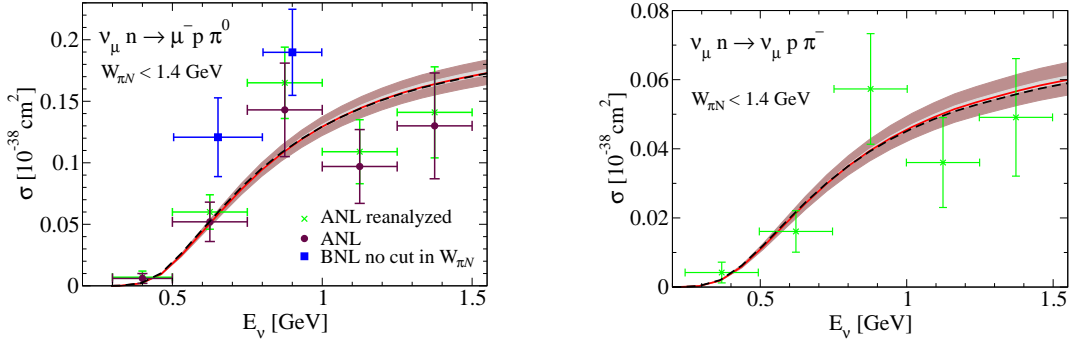

\vspace{.5cm}
\resizebox{!}{4.5cm}{\includegraphics{ppi0_juan2.eps}}\hspace*{1.35cm}
\resizebox{!}{4.5cm}{\includegraphics{ppim_juan2.eps}}
\caption{ Total $\nu_\mu n\to\mu^-p\pi^0$ (left) and $\nu_\mu n\to\nu
  p\pi^-$ (right) cross sections.  Red solid and black dashed lines
  show the results obtained in this work and those derived from fit B
  of Ref.~\cite{Alvarez-Ruso:2015eva}, respectively.  Experimental
  cross sections in the left panel have been taken from
  Ref.~\cite{Radecky:1981fn} (ANL) , Ref.~\cite{Kitagaki:1986ct} (BNL),
  and Ref.~\cite{Rodrigues:2016xjj} (ANL reanalyzed), while in the
  right panel the data have been taken from
  Ref.~\cite{Derrick:1980nr} (ANL).  Theoretical results, ANL and ANL
  reanalyzed cross sections include a $W_{\pi N} <1.4\,$GeV cut in the
  final pion-nucleon invariant mass. Experimental errors and
  theoretical bands have been evaluated as described in Fig. \ref{fig:nuevofit}. ANL reanalyzed data
  have no systematic errors due to flux uncertainties. Deuteron
  effects have been taken into account as explained in
  Ref.~\cite{Hernandez:2010bx}. }
\label{fig:otros_canales}
\end{figure}

In the lower panel,  we show the   
 $\nu_\mu n\to \mu^- n\pi^+$ cross section.   The new theoretical results  are 
 very different from the ones obtained  from fit B in 
 Ref.~\cite{Alvarez-Ruso:2015eva}, and they are now  
  in a much better global agreement with experimental data. The modifications introduced
  in the $\Delta$ contributions, that amount to the introduction of new contact
   terms controlled by the fitted LEC $c$, are crucial for this. Without those, one can not reproduce the 
$\nu_\mu n\to \mu^- n\pi^+$ cross sections without worsening the
   agreement with data in other channels.

Results for the total $\nu_\mu n\to\mu^-p\pi^0$  and $\nu_\mu n\to\nu_\mu p\pi^-$
cross sections are given in Fig.~\ref{fig:otros_canales}. We find a good 
global agreement with data and only very minor  differences with  the results
obtained from fit B carried out in Ref.~\cite{Alvarez-Ruso:2015eva}. 

The brown and gray theoretical bands in Figs.~\ref{fig:nuevofit} and \ref{fig:otros_canales}
show the sensitivity of the predicted cross sections 
to the errors  on  the the best-fit parameters $C_5^A(0)$ and the LEC
$c$, respectively. In this latter case, only the
 $\nu_\mu n\to \mu^- n\pi^+$ channel is strongly affected when varying $c$. This was
not unexpected, since the $\nu_\mu n\to \mu^- n\pi^+$ cross section has a large
contribution form the C$\Delta$P amplitude, and thus, it is very sensitive to the 
spin 1/2 part of the
$\Delta$ propagator, which strength is now controlled by the parameter $c$.

\begin{figure}
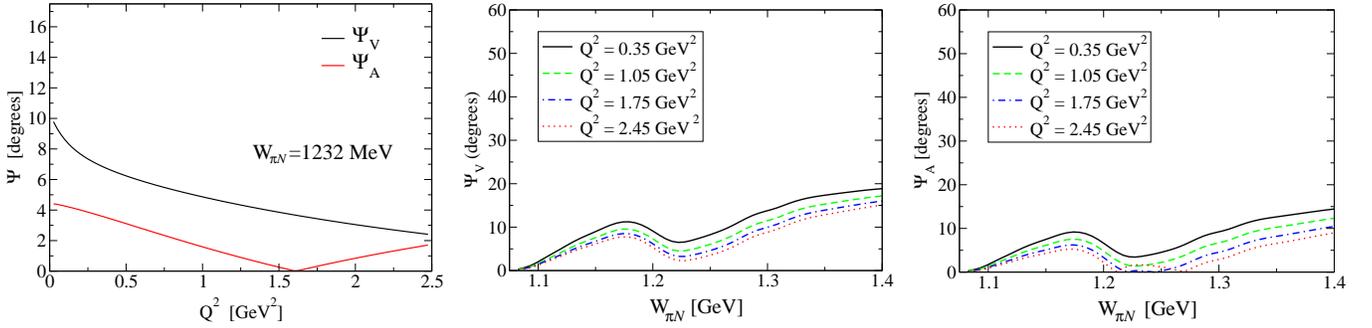

\vspace*{1cm}
\resizebox{!}{4.25cm}{\includegraphics{fases1.eps}}\hspace*{.15cm}
\resizebox{!}{4.25cm}{\includegraphics{vector.eps}}\hspace*{.15cm}
\resizebox{!}{4.25cm}{\includegraphics{axial.eps}}\caption{ Olsson phases
from the  fit carried out in this work. Left panel: 
$\Psi_V$ and $ \Psi_A$ at
the $\Delta$ peak as a function of $Q^2=-q^2$. 
Middle and right
panels: $\Psi_V$ and $ \Psi_A$ as a function of 
 the $\Delta$ invariant mass $W_{\pi N}$ for different $Q^2$ values, 
 respectively.}
  \label{fig:fases}
\end{figure}

The Olsson phases needed to satisfy Watson's theorem are presented 
 in Fig.~\ref{fig:fases}. We have selected the scales in order to allow a direct
 comparison with those obtained in 
 Ref.~\cite{Alvarez-Ruso:2015eva}, which are shown in
 Fig.~3 of that reference. We now find much smaller values, always
 below 20$^o$, and at the $\Delta$ peak (left panel in
 Fig.~\ref{fig:fases})  axial (vector) phases remain quite small and
 below 5$^o$  (10$^o$) for the whole range $\big([0,2,5]$ GeV$^2\big)$
 of $Q^2$ values shown in  the plot. This means that the present model 
 without the phases is
closer to  satisfying  unitarity than the one in
Ref.~\cite{Alvarez-Ruso:2015eva}. 

Finally, we pay attention to 
 the best-fit value quoted in Eq.~\eqref{eq:fit-results} for the LEC
 $c$. It is compatible with $-1$, within errors, but however we should
 point out that $c=-1$ does not correspond exactly to a consistent
 coupling. This is because of the $\Delta$ width, and thus even for
 $c=-1$, we have
\bea
\frac{P_{\mu\nu}}
{p_\Delta^2-M_\Delta^2+iM_\Delta\Gamma_\Delta}-\delta
P_{\mu\nu}(p_\Delta) &=& \frac{P_{\mu\nu}-
\left(p_\Delta^2-M_\Delta^2+iM_\Delta\Gamma_\Delta\right)\delta
P_{\mu\nu}(p_\Delta) }
{p_\Delta^2-M_\Delta^2+iM_\Delta\Gamma_\Delta}\nonumber \\
&=& \frac{P_{\mu\nu}-
\frac{p_\Delta^2-M_\Delta^2+iM_\Delta\Gamma_\Delta}{p_\Delta^2-M_\Delta^2}\left(
P_{\mu\nu} -
\frac{p^2_\Delta}{M_\Delta^2}P_{\mu\nu}^{\frac32}\right) }
{p_\Delta^2-M_\Delta^2+iM_\Delta\Gamma_\Delta}\nonumber \\
&=& \frac{p^2_\Delta}{M_\Delta^2} \frac{P_{\mu\nu}^{\frac32}}
{p_\Delta^2-M_\Delta^2+iM_\Delta\Gamma_\Delta}-
\frac{iM_\Delta\Gamma_\Delta}{p_\Delta^2-M_\Delta^2} \frac{P_{\mu\nu} -
\frac{p^2_\Delta}{M_\Delta^2}P_{\mu\nu}^{\frac32}}
{p_\Delta^2-M_\Delta^2+iM_\Delta\Gamma_\Delta}.\label{eq:cmenos1}
\eea
The first term  in Eq.~\eqref{eq:cmenos1} corresponds to the prescription for consistent
interactions advocated in
Refs.~\cite{Pascalutsa:1998pw,Pascalutsa:1999zz,Pascalutsa:2000kd}. The
second one, that vanishes for the C$\Delta $P amplitude, provides
complex corrections to the direct $\Delta$ contribution, which induce changes
in the Olsson phases. Indeed, we have checked that if the
second term in Eq.~\eqref{eq:cmenos1} is neglected, one finds also 
an improved description 
of the $\nu_\mu n\to \mu^- n\pi^+$ data, as compared to the $c=0$
case, and just a bit worse than that
presented here in Fig.~\ref{fig:nuevofit}. However, the needed Olsson
phases turn out to be larger than those depicted  in
Fig.~\ref{fig:fases}, being only slightly different to the ones found
in Ref.~\cite{Alvarez-Ruso:2015eva}, where the LEC $c$ was set to zero.
 Note that the $p^2_\Delta/M_\Delta^2$ factor, in front of the first
term of Eq.~\eqref{eq:cmenos1}, drastically suppresses the  C$\Delta $P
contribution,  because in this mechanism the $\Delta$
is largely off shell, with $p^2_\Delta$ much smaller (in modulus) than
$M_\Delta^2$.

\begin{figure}[h!]
\vspace{.5cm}
\resizebox{!}{6.5cm}{\includegraphics{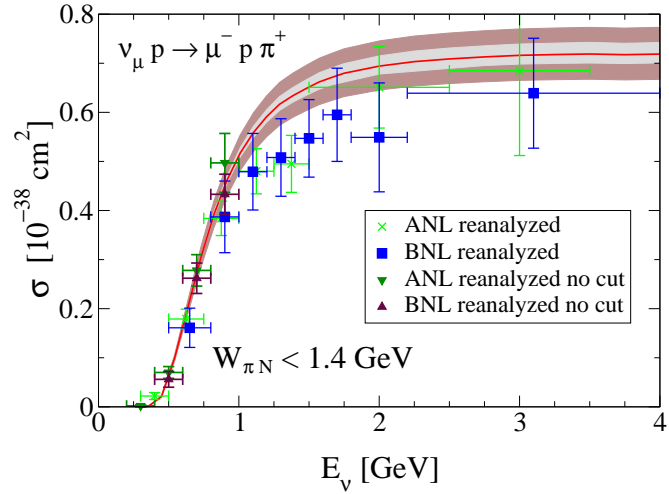}}
\caption{ Total $\nu_\mu p\to \mu^- p\pi^+$ cross section, evaluated
  with the parameters of  Eq.~\eqref{eq:fit-results} and with the $W_{\pi
    N}<1.4\,$GeV cut, compared to ANL and BNL reanalyzed data taken
  from Ref.~\cite{Rodrigues:2016xjj}. For $E_\nu<1\,$GeV, we also show
  ANL and BNL reanalyzed data where no cut in $W_{\pi N}$ has been
  applied. Theoretical bands as in Fig.~\ref{fig:nuevofit}. Deuteron
  effects have been taken into account as explained in
  Ref.~\cite{Hernandez:2010bx}.}
\label{fig:ppipcut}
\end{figure}

If one looks in more detail at the results for the $\nu_\mu n\to \mu^-
n\pi^+$ cross section shown in the lower panel of
Fig.~\ref{fig:nuevofit}, one sees that, though we obtain a global good
agreement, the model underestimates the experimental (central) values below
0.9\,GeV, while for higher energies it   overestimates the
data. This is also true, with some exceptions, for the $\nu_\mu
n\to\mu^-p\pi^0$ and $\nu_\mu n\to\nu_\mu p\pi^-$ channels depicted in
Fig.~\ref{fig:otros_canales}. In fact, the model fails to provide a
reasonable description of the central values below 1\,GeV for those
channels, using realistic values of the fitted parameters as we will
see later. This is the reason why to better determine the parameter
$c$, we included $\nu_\mu n\to \mu^- n\pi^+$ data above 1\,GeV, and
then we had to implement the cut in $W_{\pi N}$. The situation is
different for the $\nu_\mu p\to \mu^- p\pi^+$ case, where we provide a
good reproduction of the data for neutrino energies below 1\,GeV. One
might, however, look at the predictions of the model for the $\nu_\mu
p\to \mu^- p\pi^+$ cross sections, with the $W_{\pi N}<1.4\,$GeV cut,
at higher energies. The comparison of the model results with  data,
up to $4\,$GeV for the neutrino energy, is now shown in
Fig.~\ref{fig:ppipcut}.
We find an overall reasonable description of the reanalyzed ANL and
BNL cross sections, though the model also 
overestimates  the central values  for neutrino energies in
the range $1$--$2$ GeV, as  it occurred in the  other channels.  

Part of this discrepancy could be perhaps accounted for by including a
phenomenological form factor to regularize the possible unphysical behavior of
the $\Delta$ tree-level amplitudes in the kinematic regions far from
the peak of the resonance~\cite{Gonzalez-Jimenez:2016qqq}. The effects
of such form factor, with the form and parameters used in
\cite{Gonzalez-Jimenez:2016qqq}, on the $W_{\pi N}< 1.4\,$GeV $\nu_\mu
p\to \mu^- p\pi^+$ cross sections could be seen in Fig. 18 of this
latter reference, and they would certainly improve the description
exhibited in Fig.~\ref{fig:ppipcut}.  This would also improve our
reproduction of the $\nu_\mu n\to \mu^- n\pi^+$ and $\nu_\mu n\to
\mu^- p\pi^0$ data at energies above 1\,GeV. Results below 1\,GeV will
be affected to a much lesser extend, while the effects on the $\nu_\mu
p\to \mu^- p\pi^+$ flux-folded $d\sigma/dQ^2$ differential cross
section could be mostly reabsorbed into the $\beta$ flux
parameter. This is certainly a topic that is worth analyzing in future
work, paying also an special attention to its possible interference/interplay
with the partial unitarization implemented in our model through
the Olsson phases~\cite{Alvarez-Ruso:2015eva}. When
considering higher neutrino energies, it would be also advisable to
study the effects produced by the assumption of the Adler's constrains on the axial
$C_4$ and $C_5$ form factors. The contributions driven by these latter
form factors are not relevant at the low $q^2$ values accessible
when the neutrino energy is below 1 GeV~\cite{Hernandez:2010bx}, but
they might need to be considered more carefully, especially in the $\nu_\mu p\to
\mu^- p\pi^+$ channel, when higher neutrino energies are
examined. 

Nevertheless, we have also carried out a best fit taking into account
the $W_{\pi N}<1.4\,$GeV $\nu_\mu p\to \mu^- p\pi^+$ cross sections
depicted in Fig.~\ref{fig:ppipcut} instead of those below 1\,GeV,
shown in the upper right panel of Fig.~\ref{fig:nuevofit}. The new
best fit parameters differ from those quoted in
Eq.~\eqref{eq:fit-results} in one ($M_{A\Delta}$, $c$) or two
($C^A_5(0),\beta$) sigmas\footnote{The largest changes occur for
  $C^A_5(0)$ and $\beta$. The first of these parameters now takes
  values of around 1.07 leading to smaller cross sections. This needs
  to be compensated by a change of 15\% in the normalization parameter
  $\beta$, which is now around $\sim 1.05$, to avoid spoiling the
  description of the flux averaged $d\sigma/dQ^2$ differential
  $\nu_\mu p\to \mu^- p\pi^+$ cross section included in the fit.}. The
new fit is, in our view,  somehow unsatisfactory, because the
resulting model appreciably underestimates the $\nu_\mu p\to \mu^-
p\pi^+$ cross section data obtained  at 0.7 and 0.9
GeV when no cut on $W_{\pi N}$ is imposed. It is, however, precisely
at these low energies where the model, inspired in a chiral expansion,
should perform best. A word of caution must be said here. For neutrino
energies below 1\,GeV, the $W_{\pi N}<1.4\,$GeV cut does not lead to
appreciable effects on the  cross sections obtained within our
model.  This is in accordance with the data shown in Table III of
Ref.~\cite{Radecky:1981fn} (ANL), where up to 1\,GeV, there is almost
no difference between data reported with and without the cut. This
should be expected, since below 1\,GeV there is little phase space
available for $W_{\pi N}>1.4\,$GeV. However, as seen in
Fig.~\ref{fig:ppipcut}, both ANL and BNL reanalyzed data for 0.9 GeV
are significantly smaller when the cut $W_{\pi N}<1.4\,$GeV is taken
into account. Thus, it seems to be a certain degree of inconsistency
between the two $\nu_\mu p\to \mu^- p\pi^+$ data sets (with or without
the $W_{\pi N}<1.4\,$GeV cut) below 1\,GeV. As a result, we can fit
the parameters in our model to reproduce one or the other set of cross
sections, but not both at the same time. We preferred to use the
reanalyzed data without the $W_{\pi N}<1.4\,$GeV cut since their
extraction seem to suffer from less uncertainties\footnote{As stated
  in Ref.~\cite{Rodrigues:2016xjj}, to get the reanalyzed $W_{\pi
    N}<1.4\,$GeV cross sections, the ratio of reanalyzed to published
  cross sections obtained without the cut is used.}.

Finally, we have also explored the possibility of fitting only data
below 1\, GeV and with no $W_{\pi
    N}<1.4\,$GeV cut applied. To that end, we have included in the fit
 $\nu_\mu p\to \mu^- p\pi^+$ , $\nu_\mu n\to \mu^- n\pi^+$ and 
   $\nu_\mu
n\to \mu^- p\pi^0$ data  in this neutrino energy range taken from 
Ref.~\cite{Rodrigues:2016xjj}. In
this new fit, the $c$ parameter significantly departs from $-1$
(propagation of only spin 3/2 degrees of freedom in the C$\Delta $P
term) and becomes closer to $-1.5$, while $C_5^A(0)$ is about 1.23,
now even above the GTR prediction. However the normalization parameter
$\beta$ turns out to be 1.35, a value too large to be accommodated
within the ANL flux uncertainties. Besides, one obtains 
$\chi^2/dof=3.07$, which is much worse than for our preferred fit in 
Eq.~(\ref{eq:fit-results}).

Thus, we consider the fit of Eq.~\eqref{eq:fit-results} a sensible
option, given the somehow uncertain situation, and that it leads to a
remarkable description of the pion photoproduction data off the
nucleon, as we will discuss next.

\subsection{Pion photoproduction}
\begin{figure}[h!]
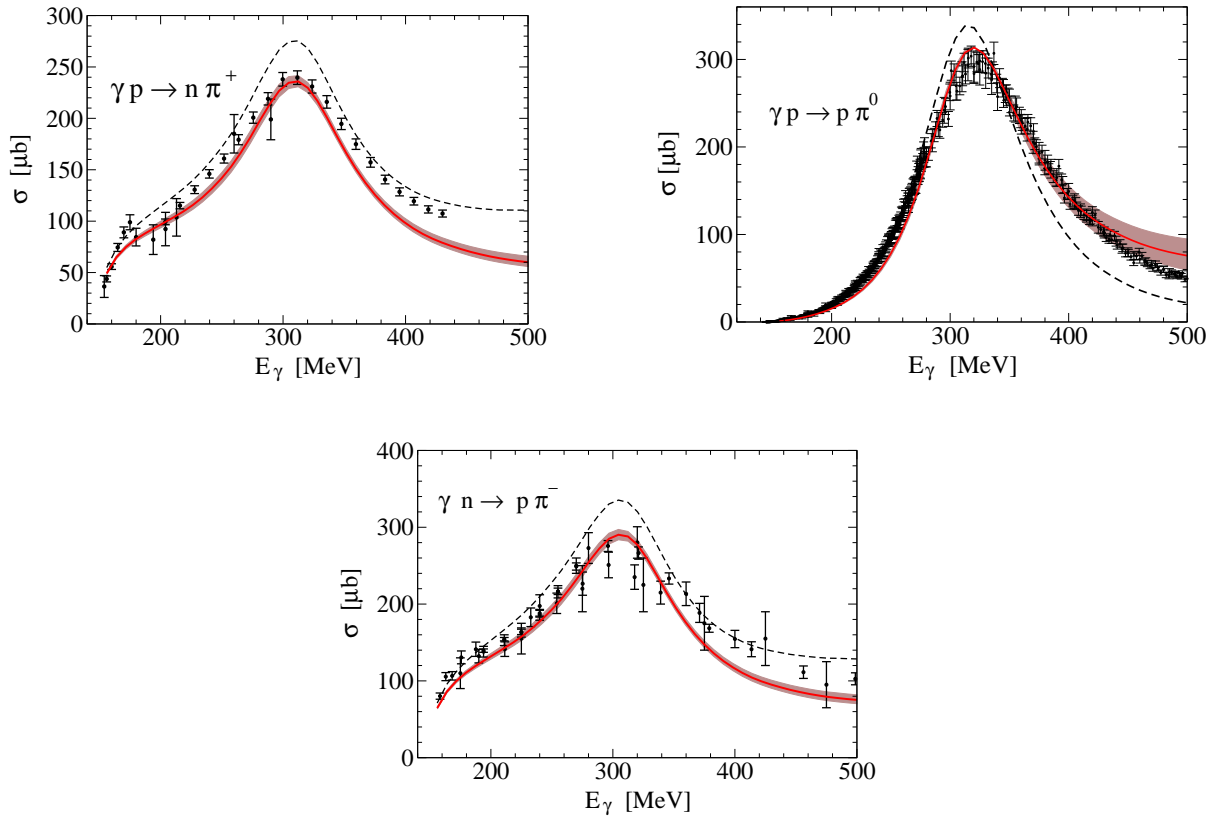

\vspace{.5cm}
\resizebox{!}{5cm}{\includegraphics{npip_em_juan2.eps}}\hspace*{1.35cm}
\resizebox{!}{5cm}{\includegraphics{ppi0_em_juan2.eps}}\vspace{.75cm}\\
\resizebox{!}{5cm}{\includegraphics{ppim_em_juan2.eps}}
\caption{ Total  $\gamma p\to n\pi^+$ (upper left), $\gamma p\to p\pi^0$ 
(upper right) and
$\gamma n\to p\pi^-$ (bottom) cross sections as a function of the
  photon energy in the laboratory frame.  Red solid and black
  dashed lines show the  predictions from the model presented in this work (see 
the Appendix) and the
results obtained without the  modification  of the spin 1/2 component of the 
$\Delta$ propagator ($c=0$). Cross sections have been taken from the George Washington University SAID
database~\cite{SAID}. Theoretical uncertainty bands
account for the variation of the results when the parameter $c$ changes
within its error interval given in Eq.~\eqref{eq:fit-results}.}
\label{fig:em}
\end{figure}

Since in Ref.~\cite{Alvarez-Ruso:2015eva}, we showed results for pion
photoproduction, it is relevant to see also for this case the effect
of the modification introduced in the $\Delta$ propagator. Amplitudes
for pion photoproduction derive directly from the vector part of our
model for weak pion production by neutrinos and they are extensively
discussed in the Appendix. As for the case of neutrino production, the
model is also partially unitarized by imposing Watson's theorem on the
dominant vector multipole, now evaluated at $q^2=0$.  What we will show are  pure
predictions of the model without any readjustment of parameters or
vector form factors. In Fig.~\ref{fig:em} we present results for total
cross sections that we compare to data taken from the George
Washington University SAID database~\cite{SAID}. On the theoretical
side, we compare the predictions obtained with the present model (red
solid lines) with the results obtained without the modification of the
spin 1/2 component of the $\Delta$ propagator (black dashed lines),
the latter corresponding to setting $c=0$. The description of the data
is better in the current modified case, with $c$ close to $-1$.  The theoretical bands show
the sensitivity of the results with respect to the $c$ parameter, when
it is varied within the errors quoted in
Eq.~\eqref{eq:fit-results}. To get a better reproduction of the cross
sections above the $\Delta$ resonance region, the model would have to
be enlarged by the addition of extra resonance contributions relevant
for the case of electro- or photoproduction.

\section{Summary and conclusions}
\label{sec:conclusions}
We have improved our model of Refs.~\cite{Hernandez:2007qq,
  Hernandez:2013jka,Alvarez-Ruso:2015eva} by including two extra
contact terms\footnote{The correction in the $\Delta$
propagator of Eq.~(\ref{eq:modifipropa}) induces contact interactions both for
the $\Delta\rm P$ and $\rm C\Delta\rm  P$ amplitudes in the original model of 
Refs.~\cite{Hernandez:2007qq, Alvarez-Ruso:2015eva}.}. This has been motivated by the failure of present
theoretical approaches to describe the $\nu_\mu n\to\mu^-n\pi^+$ total
cross section data. As shown in Ref.~\cite{Hernandez:2007qq}, this
channel has a large contribution from the C$\Delta$P mechanism and it
is thus very sensitive to the spin 1/2 components in the $\Delta$
propagator. This spin 1/2 part is nonpropagating and it gives rise to
contact terms. Contact terms appear naturally within effective field
theories, and in particular in ChPT, as counterterms with unknown
strengths. Indeed, the coefficients of the contact terms have to be
ultimately fitted to experiment. Aiming at keeping our model simple,
we have just introduced only one new parameter, $c$, that controls the
strength of the contact terms generated by the spin 1/2 part of the
$\Delta$ propagator. To constraint its value, we have also included
$\nu_\mu n\to\mu^-n\pi^+$ cross section data in the fit. The description
of this channel considerably improves, without affecting the good
results we had already obtained in Refs.~\cite{Hernandez:2007qq,
  Alvarez-Ruso:2015eva} for the other channels. Since the fitted value
of $c$ is compatible with $-1$, we find that the  crossed $\Delta$ 
pole amplitude is substantially suppressed and that consistent
$\Delta$
couplings~\cite{Pascalutsa:1998pw,Pascalutsa:1999zz,Pascalutsa:2000kd}
are preferred. 
Besides, the new Olsson phases
needed to satisfy Watson's theorem are now much smaller than those
obtained in Ref.~\cite{Alvarez-Ruso:2015eva} for the $c=0$ case,
indicating that the present version without the phases is closer to
satisfying unitarity. Yet, the  $C_5^A(0)$ is now larger by 3.5\% than that found in 
Ref.~\cite{Alvarez-Ruso:2015eva}, and it is in remarkable agreement
with the GTR prediction. 
  
  We have also explored how this change in the $\Delta$ propagator 
  affects our predictions for pion
  photoproduction.  We also find now a better agreement with experiment  compared
  to the case where the LEC $c$ was set to zero.

Finally, we should mention that  FSI effects on single pion production off the deuteron might
induce corrections on the nucleon spectator approximation. This
approximation is used to extract the pion production cross sections on
the nucleon from the data on the deuteron. These effects have not been
addressed in this work. However, it has been argued~\cite{Wu:2014rga,
  Nakamura:2016cnn} that they might be of special relevance precisely
in the $n\pi^+$ channel, and that the ANL and BNL data on the
deuterium target might need a more careful analysis with the FSI's
taken into account. For such a reanalysis to be meaningful, it will be
mandatory to incorporate the kinematical cuts implemented in the old
experiments to properly separate the three reaction channels ($p\pi^+,
p\pi^0$, and $n\pi^+$), since these cuts were designed to minimize the
corrections to the spectator hypothesis. Nevertheless, the existence
of some FSI effects will not exclude the solution to the $n\pi^+$
puzzle offered here, and based on the possibility of adding
phenomenological contact terms. It is certainly natural within the
context of effective field theories.

%
%
%
%
%
\begin{acknowledgments}
We acknowledge discussions with A. Mariano and A. Pich. This research
 has been supported by the Spanish Ministerio de Econom\'\i a y
 Competitividad and European FEDER funds under  Contracts No.
 FPA2013-47443-C2-2-P, No. FIS2014-51948-C2-1-P,  No. FIS2014-57026-REDT. and 
 No. SEV-2014-0398, and by
 Generalitat Valenciana under Contract No. PROMETEOII/2014/0068.
 \end{acknowledgments}
%
%
%
%
%

\appendix

\section{Model for pion photo- and electroproduction off the nucleon}
\label{app:model_em}
\begin{figure}[tbh]
\resizebox{!}{8cm}{\includegraphics{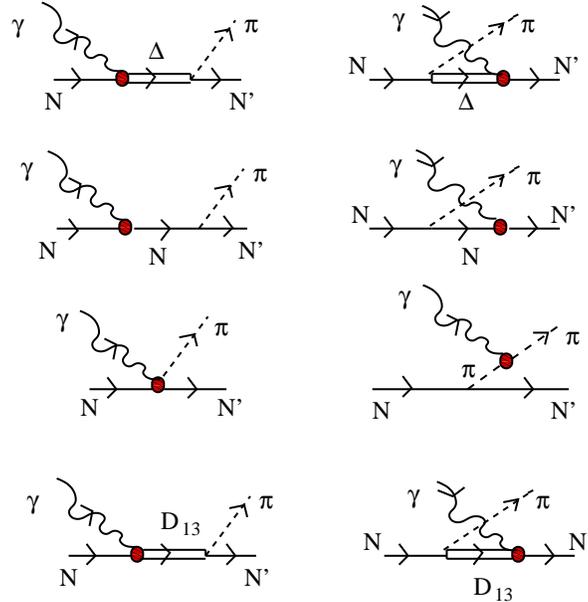}}
\caption{ Model for the $\gamma N\to N^\prime\pi$ or $\gamma^* N\to N^\prime\pi$ 
reactions. First row:  Direct and crossed
  $\Delta(1232)$  pole terms. Second row: Direct and crossed nucleon
    pole terms. Third row: Contact and   pion-in-flight terms. Fourth row:
    Direct and crossed $D_{13}$ pole terms. }
  \label{fig:diagramas_em}
\end{figure}

Our model for pion photo or electroproduction off the nucleon derives
directly from the vector part of that constructed for weak pion
production by neutrinos. Thus, it includes all the contributions
depicted in Fig.~\ref{fig:diagramas_em}: the resonant direct and
crossed $\Delta(1232)$ pole terms ($\Delta$P and C$\Delta$P,
respectively) and the background terms required by chiral symmetry.  The
latter ones include direct and crossed nucleon pole (labeled as NP
and CNP), contact  (CT) and pion-in-flight (PF) terms. Besides we
also consider the direct and crossed $D_{13}(1520) $ pole terms (DP and CDP,
respectively).

In the notation of Ref.~\cite{Hernandez:2007qq}, the quark level  electromagnetic current is given by\footnote{We ignore
the contribution from heavy quarks.}
\begin{eqnarray}
s^\mu_{\rm em} &=& \frac23 \bar{\Psi}_u\gamma^\mu \Psi_u 
- \frac13 \bar{\Psi}_d\gamma^\mu \Psi_d 
- \frac13 \bar{\Psi}_s\gamma^\mu \Psi_s.
\eea
This can be written as the sum of an isoscalar and an isovector pieces
\bea
s^\mu_{\rm em} &=& s^\mu_{\rm em\ IS} + s^\mu_{\rm em\ IV}\\
s^\mu_{\rm em\ IS}&=& \frac16 
\bar \Psi_q\gamma^\mu\Psi_q - \frac13 \bar\Psi_s \gamma^\mu\Psi_s,\nonumber\\
s^\mu_{\rm em\ IV}&=&
\frac{1}{\sqrt  2} \bar\Psi_q\,\gamma^\mu \frac{\tau^1_0}{\sqrt 2} \,\Psi_q
\end{eqnarray}
with $\Psi_q=\left(\begin{array}{c}\Psi_u\\\Psi_d\end{array}\right)$ and 
$\tau^1_0=\tau_z$, where $\tau_x,\tau_y,\tau_z$ are the three Pauli matrices.

In the same notation the vector part of the charged weak current reads
\bea
V^\mu_{\rm cc\pm}=\mp\bar\Psi_q\,\gamma^\mu
\frac{\tau^1_{\pm1}}{\sqrt 2} \,\Psi_q
\eea
with $\tau^1_{\pm1}=\mp\frac1{\sqrt2}(\tau_x\pm i\tau_y)$. We could
relate the matrix elements of the isovector part 
of the electromagnetic current with those of  $V^\mu_{\rm cc\pm}$. To
that end, we express the physical nucleon-pion states in terms of
states  with well-defined total isospin,
\bea
\label{eq:isospin}
|p\pi^+\rangle&=&-|N\pi; 3/2,3/2\rangle,\nonumber\\
|p\pi^0\rangle&=&\sqrt{\frac23}|N\pi; 3/2,1/2\rangle+
\frac1{\sqrt3}|N\pi; 1/2,1/2\rangle,\nonumber\\
|n\pi^+\rangle&=&-\frac{1}{\sqrt3}|N\pi; 3/2,1/2\rangle+
\sqrt{\frac23}|N\pi; 1/2,1/2\rangle,\nonumber\\
|n\pi^0\rangle&=&\sqrt{\frac23}|N\pi; 3/2,-1/2\rangle-
\frac1{\sqrt3}|N\pi; 1/2,-1/2\rangle,\nonumber\\
|p\pi^-\rangle&=&\frac{1}{\sqrt3}|N\pi; 3/2,-1/2\rangle+
\sqrt{\frac23}|N\pi; 1/2,-1/2\rangle,\nonumber\\
|n\pi^-\rangle&=&|N\pi; 3/2,-3/2\rangle.
\eea  
and then we can obtain\footnote{ For a tensor
 operator $T^j_m$, we use the
 Wigner-Eckart theorem with the convention
 \bea
 \langle j_2m_2|T^j_m|j_1m_1\rangle=(j_1,j,j_2,m_1,m,m_2)\,
 \langle j_2\parallel
 T^j\parallel j_1\rangle,
 \eea
 with $(j_1,j,j_2,m_1,m,m_2)$ a Clebsch-Gordan coefficient and 
 $\langle j_2\parallel
 T^j\parallel j_1\rangle$ the reduced matrix element.}
\bea
\langle p\pi^+|V^\mu_{cc+}(0)|p\rangle&=&-\langle 3/2\parallel
V^\mu\parallel1/2\rangle,\nonumber\\
\langle n\pi^+|V^\mu_{cc+}(0)|n\rangle&=&-\frac{1}{\sqrt3}\langle N\pi; 3/2,1/2|
V^\mu_{cc+}(0)|n\rangle+
\sqrt{\frac23}\langle N\pi;
1/2,1/2|V^\mu_{cc+}(0)|n\rangle\nonumber\\
&=&-\frac13\langle 3/2\parallel
V^\mu\parallel1/2\rangle-\frac23\langle 1/2\parallel
V^\mu\parallel1/2\rangle,
\eea 
from where
\bea
\langle 3/2\parallel
V^\mu\parallel1/2\rangle&=&-\langle p\pi^+|V^\mu_{cc+}(0)|p\rangle,\nonumber\\
\langle 1/2\parallel
V^\mu\parallel1/2\rangle&=&-\frac32\langle
n\pi^+|V^\mu_{cc+}(0)|n\rangle+\frac12\langle p\pi^+|V^\mu_{cc+}(0)|p\rangle.
\eea
These two reduced matrix elements determine all matrix elements 
of the isovector part of the electromagnetic current. As an example, 
we evaluate\footnote{Note the factor $-\frac1{\sqrt2}$ difference in the
definition of $s^\mu_{\rm em\ IV}$ and $V^\mu_{cc+}$.}
\bea
\langle p\pi^0|s^\mu_{\rm em\ IV}(0)|p\rangle&=&\sqrt{\frac23}
\langle N\pi; 3/2,1/2|s^\mu_{\rm em\ IV}(0)|p\rangle+
\frac1{\sqrt3}\langle N\pi; 1/2,1/2|s^\mu_{\rm em\ IV}(0)|p\rangle\nonumber\\
&=&-\frac1{\sqrt2}\Big(\frac23\langle 3/2\parallel
V^\mu\parallel1/2\rangle+\frac13\langle 1/2\parallel
V^\mu\parallel1/2\rangle\Big)\nonumber\\
&=&\frac1{2\sqrt2}\left(\,\langle p\pi^+|V^\mu_{cc+}(0)|p\rangle+\langle n\pi^+|
V^\mu_{cc+}(0)|n\rangle\,\right).
\eea
Similarly,
\bea
\langle n\pi^+|s^\mu_{\rm em\ IV}(0)|p\rangle&=&-\frac12
\left(\,\langle p\pi^+|V^\mu_{cc+}(0)|p\rangle-\langle n\pi^+|
V^\mu_{cc+}(0)|n\rangle\,\right),\nonumber\\
\langle n\pi^0|s^\mu_{\rm em\ IV}(0)|n\rangle&=&\langle p\pi^0|s^\mu_{\rm em\ IV}(0)|p\rangle\nonumber\\
\langle p\pi^-|s^\mu_{\rm em\ IV}(0)|n\rangle&=&- \langle n\pi^+|s^\mu_{\rm em\ IV}(0)|p\rangle.
\eea

Since the $\Delta$ exchange contributions of Fig~\ref{fig:diagramas_em} 
are purely
isovector, and denoting by $j^\mu_{\rm em}$ the matrix elements of the
electromagnetic current, 
we thus get\footnote{The Feynman amplitude will be proportional to
  $e j^\mu_{\rm em}\,\epsilon_\mu$, with $\epsilon_\mu$ the photon
  polarization vector and $e=\sqrt{4\pi\alpha}$, the dimensionless
  proton electric charge, with $\alpha\sim 1/137$.}
\bea
j^\mu_{\rm em}|_{\Delta \rm P} &=& {\rm
  i}\,C^{\Delta\rm P}_\gamma\frac{f^*}{m_\pi}\sqrt{3}\,
  {k_\pi^\alpha}
  \bar u(\vec{p}\,')
  \left[\frac{P_{\alpha\beta}(p_\Delta)}
  {p_\Delta^2-M_\Delta^2+ i M_\Delta \Gamma_\Delta}
  +c\,\delta P_{\alpha\beta}(p_\Delta)
  \right]
  \Gamma^{\beta\mu}_V(p,q)
  u(\vec{p}\,),\nonumber \\\nonumber \\ \quad p_\Delta&=&p+q, \ \  
 C^{\Delta\rm P}_\gamma = \left( \begin{array}{ccc} 
\sqrt2/3   & {\rm for} & p\to p\pi^0 \\ 
-1/3   & {\rm for} & p\to n\pi^+ \\  
  \sqrt2/3
  & {\rm for} & n\to n\pi^0  \\
  1/3
  & {\rm for} & n\to p\pi^- \end{array} \right ),\nonumber\\
  \Gamma_{V}^{\beta\mu} (p,q) &=&\left [ \frac{C_3^V}{M}\left(g^{\beta\mu} 
  \slashchar{q}-
q^\beta\gamma^\mu\right) + \frac{C_4^V}{M^2} \left(g^{\beta\mu}
q\cdot p_\Delta- q^\beta p_\Delta^\mu\right)
+ \frac{C_5^V}{M^2} \left(g^{\beta\mu}
q\cdot p- q^\beta p^\mu\right) + C_6^V g^{\beta\mu}
\right ]\gamma_5,\quad p_\Delta= p+q  \nonumber\\
  \eea
\bea
j^\mu_{\rm em}|_{\rm C\Delta\rm P} &=& {\rm
  i}\,C^{\rm C\Delta\rm  P}_\gamma\frac{f^*}{m_\pi}\frac{1}{\sqrt 3}
  \,{k_\pi^\beta}
  \bar u(\vec{p}\,')  {\widehat \Gamma}^{\mu\alpha}_V(p',q)
   \left[\frac{P_{\alpha\beta}(p_\Delta)} 
   {p_\Delta^2-M_\Delta^2+ i M_\Delta \Gamma_\Delta}
  +c\,\delta P_{\alpha\beta}(p_\Delta)
  \right] 
  u(\vec{p}\,),\nonumber \\\nonumber \\\quad p_\Delta&=&p'-q, \ \ 
 C^{\rm C\Delta\rm  P}_\gamma = \left( \begin{array}{ccc} 
\sqrt2 & {\rm for} & p\to p\pi^0 \\
1 & {\rm for} & p\to n\pi^+ \\
 \sqrt2   & {\rm for} & n\to n\pi^0\\
 -1& {\rm for} & n\to p\pi^-\\
  \end{array} \right ), \quad {\widehat \Gamma}^{\mu\alpha}_V(p',q) = 
  \gamma^0\left
  [\Gamma^{\alpha\mu}_V(p', -q)\right]^\dagger \gamma^0  
  \end{eqnarray}
where $q$, $p$, $k_\pi$,  and $p'$ are the incoming photon and nucleon
and the outgoing pion and nucleon four momenta.

To compute the nonresonant amplitudes, we pay attention to the electromagnetic current
associated to the Lagrangian of the SU(2) nonlinear $\sigma$ model
derived in Ref.~\cite{Hernandez:2007qq}. It reads,
\begin{equation}
s^\mu_{\rm em} = \bar\Psi \gamma^\mu \left ( \frac{1+\tau_z}{2}\right) \Psi
+ \frac{{\rm i}g_A}{2f_\pi} \bar\Psi \gamma^\mu \gamma_5 \left (
\tau^1_{-1} \phi^\dagger + \tau^1_{+1} \phi \right ) \Psi + {\rm i}
\left (\phi^\dagger \partial^\mu \phi -  \phi \partial^\mu
\phi^\dagger \right) + \cdots \label{eq:real_sem}
\end{equation}
with $g_A=1.26$, $f_\pi=93.2\,$MeV,
$\Psi$ 
 and $\vec \phi$ the nucleon and
  pion fields  already introduced in
  Sec.~\ref{sec:piNDelta}. We have only kept those terms
  contributing to one pion production in the absence of chiral loop
  corrections. Thus, within our framework, and besides the excitation
  of the $\Delta$ and the $N^*(1520)$, the model for the $\gamma N \to
  \pi N$ reaction would consist of direct and crossed nucleon pole,
  contact and pion-in-flight terms, as shown diagrammatically
  in  Fig.~\ref{fig:diagramas_em}. We see that neither the pion-in-flight nor the contact terms contribute for
$\pi^0$ photoproduction, which implies in turn that they are purely
  isovector. Thus, we get for these two contributions
\begin{eqnarray}
j^\mu_{\rm em}|_{\rm CT} &=& 
-{\rm i}\,C^{\rm CT}_\gamma\frac{g_A}{\sqrt 2 f_\pi}\left(F_{1}^p(q^2)-F_{1}^n(q^2)
\right)\
  \bar u(\vec{p}\,') \gamma^\mu
  \gamma_5  u(\vec{p}\,),\quad C^{\rm CT}_\gamma = \left( \begin{array}{ccc} -1
  & {\rm for} & p\to n\pi^+ \cr \phantom{-}1
  & {\rm for} & n\to p\pi^- \end{array} \right ) \\
j^\mu_{\rm em}|_{\rm PF} &=& 
-{\rm i}\,C^{\rm PF}_\gamma\frac{g_A}{\sqrt 2 f_\pi}\left(F_{1}^p(q^2)-F_{1}^n(q^2)
\right)  \
  \frac{2M (2k_\pi-q)^\mu}{(k_\pi-q)^2-m_\pi^2}
  \bar u(\vec{p}\,')  \gamma_5 u(\vec{p}\,),\quad C^{\rm PF}_\gamma = 
  \left( \begin{array}{ccc} -1
  & {\rm for} & p\to n\pi^+ \cr \phantom{-}1
  & {\rm for} & n\to p\pi^- \end{array} \right )
\end{eqnarray}
%
For the proton and neutron Dirac electromagnetic form factors, $F_{1}^{p,n}$ we use the 
parametrization of  Galster et 
al.~\cite{Galster:1971kv}, as we did  in Ref.~\cite{Hernandez:2007qq}
for weak pion production. 

To account for direct and crossed nucleon pole contributions, we need
to consider, in addition to the isovector part, the isoscalar part of
the electromagnetic current. For the isoscalar part of the
electromagnetic current we have from Eq.~(\ref{eq:isospin})
\begin{eqnarray}
\langle n \pi^+ \big|  s^\mu_{{\rm em}\ IS} \big | p \rangle &= &\langle p \pi^- \big|
  s^\mu_{{\rm em}\ IS} \big | n \rangle =
 \sqrt 2 \langle p \pi^0
 \big|  s^\mu_{{\rm em}\ IS} \big | p
 \rangle= -\sqrt 2 \langle n \pi^0 \big|  s^\mu_{{\rm em}\ IS}
\big | n \rangle \label{eq:65}
\end{eqnarray}
Using the current of Eq.~\eqref{eq:real_sem}, supplemented
by including i) the $q^2$ dependence induced by the Dirac $F_1^{p,n}$
form factors and  ii) the magnetic contribution in the $\gamma NN$
vertex [with the corresponding magnetic form factors $\mu_p
F_2^p(q^2), \mu_n F_2^n(q^2)$,  for which we also use the Galster
parametrization], we find~\cite{Hernandez:2007qq}  
\begin{eqnarray}
\langle p \pi^0 \big|  s^\mu_{{\rm em}\ IS} \big | p \rangle &=& -  \frac{\langle
n \pi^0 | s^\mu_{\rm em}(0) | n \rangle-\langle
p \pi^0 | s^\mu_{\rm em}(0) | p \rangle }{2} \\
&=& -{\rm i}\,\frac{g_A}{
  2 f_\pi}  \bar u(\vec{p}\,') \Bigg \{
 \slashchar{k}_\pi\gamma_5\frac{\slashchar{p}+\slashchar{q}+M}{(p+q)^2-M^2+ i\epsilon}\left [F^{IS}_1(q^2)\gamma^\mu+{\rm i}\mu_{IS}\frac{F_2^{IS}(q^2)}{2M} \sigma^{\mu\nu}q_\nu 
\right]  \nonumber \\
&&+\left [F^{IS}_1(q^2)\gamma^\mu+{\rm i}\mu_{IS}\frac{F_2^{IS}(q^2)}{2M} \sigma^{\mu\nu}q_\nu 
\right] \frac{\slashchar{p}'-\slashchar{q}+M}{(p'-q)^2-M^2+ i\epsilon}
 \slashchar{k}_\pi\gamma_5 \Bigg\} u(\vec{p}\,)
\end{eqnarray}
with 
\begin{equation}
 F_1^{IS}(q^2) =  \frac12 \left (F_1^p(q^2)+F_1^n(q^2)\right),\qquad
 \mu_{IS} F_2^{IS}(q^2) = \frac12 \left ( \mu_p F_2^p(q^2) + \mu_n F_2^n(q^2)\right) 
\end{equation}
where we have made use of the cancellation of the isovector
contributions in the  difference $(\langle
n \pi^0 | s^\mu_{\rm em}(0) | n \rangle-\langle
p \pi^0 | s^\mu_{\rm em}(0) | p \rangle)$. 

Taking also into account the isovector contributions, we get the following 
direct and crossed nucleon pole amplitudes:
  %
  %
  %
  \begin{eqnarray}
j^\mu_{\rm em}|_{\rm NP} = 
-{\rm i}\,C^{\rm NP}_\gamma\frac{g_A}{ 2 f_\pi}\
  \bar u(\vec{p}\,') 
 \slashchar{k}_\pi\gamma_5\frac{\slashchar{p}+\slashchar{q}+M}{(p+q)^2-M^2+ 
 i\epsilon} V^\mu_{NP}(q)  
u(\vec{p}\,),\nonumber\\
 C^{\rm NP}_\gamma = \left( \begin{array}{ccc} 
1  & {\rm for} & p\to p\pi^0 \cr 
\sqrt2   & {\rm for} & p\to n\pi^+\cr
-1  & {\rm for} & n\to n\pi^0 \cr 
\sqrt2  & {\rm for} & n\to p\pi^-
 \end{array} \right ),\quad V^\mu_{\rm NP} = \left( \begin{array}{ccc} 
V^\mu_p(q)  & {\rm for} & p\to p\pi^0 \cr 
V^\mu_p(q)   & {\rm for} & p\to n\pi^+\cr
V^\mu_n(q)  & {\rm for} & n\to n\pi^0 \cr 
V^\mu_n(q)  & {\rm for} & n\to p\pi^-
 \end{array} \right )
\end{eqnarray}%
\begin{eqnarray}
j^\mu_{\rm em}|_{\rm CNP} = 
-{\rm i}\,C^{\rm CNP}_\gamma\frac{g_A}{ 2 f_\pi}\
  \bar u(\vec{p}\, ') V^\mu_{CNP}(q)
\frac{\slashchar{p}'-\slashchar{q}+M}{(p'-q)^2-M^2+ i\epsilon} 
\slashchar{k}_\pi
\gamma_5  u(\vec{p}\,),\nonumber\\ C^{\rm CNP}_\gamma = \left( \begin{array}{ccc} 
1  & {\rm for} & p\to p\pi^0 \cr 
\sqrt2
  & {\rm for} & p\to n\pi^+\cr
-1  & {\rm for} & n\to n\pi^0 \cr  
  \sqrt2  & {\rm for} & n\to p\pi^-
 \end{array} \right ),\quad V^\mu_{\rm CNP} = \left( \begin{array}{ccc} 
V^\mu_p(q)  & {\rm for} & p\to p\pi^0 \cr 
V^\mu_n(q)  & {\rm for} & p\to n\pi^+\cr
V^\mu_n(q)  & {\rm for} & n\to n\pi^0 \cr 
V^\mu_p(q)   & {\rm for} & n\to p\pi^-
 \end{array} \right )
\end{eqnarray}
with  
\begin{eqnarray}
V^\mu_{p,n}(q)&=&F_{1}^{p,n}(q^2)\gamma^\mu+
i\mu_{p,n}\frac{F_{2}^{p,n}(q^2)}{2M}\sigma^{\mu\nu}q_\nu
\end{eqnarray}
One can check that CVC is preserved by the nonresonant amplitudes.

Finally, we give the expressions for the DP and CDP $N^*(1520)$ terms. The isovector
parts are determined, as for the case of the $\Delta$,  in terms
of the matrix elements of the $V^\mu_{cc+}$ weak vector current that appear in 
the Appendix of Ref.~\cite{Hernandez:2013jka}. They  are given by
\begin{eqnarray} 
j^\mu_{\rm em\ IV}|_{\rm DP}&=&iC^{\rm DP}_{\rm IV}g_D\frac1{2\sqrt3}
\frac{k_\pi^\alpha}{p_D^2-M_D^2+iM_D\Gamma_D}
\bar u(\vec{p}\,' )\gamma_5\, P^D_{\alpha\beta}(p_D)
\Gamma^{V\,\beta\mu}_{D}
\left(p,q \right)
u(\vec{p}\,),\nonumber\\&&\hspace{3cm} p_D=p+q,\ \
 C^{\rm DP}_{\rm IV}=\left(\begin{array}{cll}
1&{\rm for}&p\to p\pi^0\\
\sqrt{2}&{\rm for}&p\to n\pi^+\\
1&{\rm for}&n\to n\pi^0\\
-\sqrt{2}&{\rm for}&p\to p\pi^-\\
\end{array}\right)\nonumber\\
\eea
\bea
j^\mu_{\rm em\ IV}|_{\rm CDP}&=&-iC^{\rm CDP}_{\rm IV}g_D\frac1{2\sqrt3}
\frac{k_\pi^\alpha}{p_D^2-M_D^2+iM_D\Gamma_D}
\bar u(\vec{p}\,' )\,\widehat \Gamma^{D\,\mu\beta}_{V}
\left(p',-q \right) \,
P^D_{\beta\alpha}(p_D)\, 
\gamma_5 u(\vec{p}\,),\nonumber\\
&&\hspace{0cm}p_D=p'-q,\ \ 
  C^{\rm CDP}_{\rm IV}=
\left(\begin{array}{cll}
1&{\rm for}&p\to p\pi^0\\
-\sqrt{2}&{\rm for}&p\to n\pi^+\\
1&{\rm for}&n\to n\pi^0\\
\sqrt{2}&{\rm for}&p\to p\pi^-\\
\end{array}\right),\ \ 
\widehat \Gamma^{D\,\mu\beta}_V
\left(p',-q \right)=\gamma^0[\Gamma^{D\,\beta\mu}_V
\left(p',-q \right)]^\dag\gamma^0.
\eea
with $M_D=1520$ MeV, and 
\bea
P_{\alpha\beta}^{D}(p_D)&=&-\left(\slashchar{p}_D+M_D\right)\left(g_{\alpha\beta}-\frac13\gamma_\alpha\gamma_\beta-
\frac23\frac{p_{_D\alpha}\, p_{_{D}\beta}}{M_D^2}+
\frac{1}{3}\frac{p_{_D\alpha}\,\gamma_\beta-
p_{_D\beta}\,\gamma_\alpha}{M_D}\right)\\
\Gamma^{D\,\beta\mu}_{V} (p,q) &=&\left [ \frac{\tilde C_3^V}{M}\left(g^{\beta\mu}
\slashchar{q}-
q^\beta\gamma^\mu\right) + \frac{\tilde C_4^V}{M^2} \left(g^{\beta\mu}
q\cdot p_D- q^\beta p_D^\mu\right)
+ \frac{\tilde C_5^V}{M^2} \left(g^{\beta\mu}
q\cdot p- q^\beta p^\mu\right)\right. + \tilde C_6^V g^{\beta\mu}
\bigg ],\ \ p_D=p+q.\nonumber\\
\label{eq:d13-2}
\eea
The corresponding vector form factors are given in Ref.~\cite{Hernandez:2013jka} and they are
obtained from a fit to results in Ref.~\cite{Leitner:2008wx}.\\

The value of the $g_D$ strong coupling is determined from the 
$\Gamma_{D_{13}\to N\pi}(M_D)$ partial decay width to be $g_D=20\,$GeV$^{-1}$.
This partial decay width is  given,
 for $W_{\pi N}>M+m_\pi$, by
\bea
\Gamma_{D_{13}\to N\pi}(W_{\pi N})=\frac{g_D^2}{8\pi}\frac1{3W_{\pi N}^2}
[(W_{\pi N}-M)^2-m_\pi^2]|\vec p_\pi|^3
\eea
with  $|\vec
p_\pi|=\frac{\lambda^{1/2}(W^2_{\pi N},M^2,m_\pi^2)}
{2W_{\pi N}}$. For $\Gamma_{D_{13}\to N\pi}(M_D)$ we took 61\% of 115\,MeV. For the total width
$\Gamma_D$ we use
\bea
\Gamma_D(W_{\pi N})=\Gamma_{D_{13}\to N\pi}(W_{\pi N})+
\Gamma_{D_{13}\to \Delta\pi}(W_{\pi N}).
\eea
where for  $\Gamma_{D_{13}\to\Delta\pi}$ we assumed an $S-$wave decay and took
\bea
\Gamma_{D_{13}\to\Delta\pi}(W_{\pi N})=0.39\times115\,{\rm MeV}
\frac{|\vec p_\pi^{\ \prime}|}{|\vec p_\pi^{\ \prime\,
o-s}|}\,\theta(W_{\pi N}-M_\Delta-m_\pi),
\eea
with $|\vec
p_\pi^{\ \prime}|=\frac{\lambda^{1/2}(W^2_{\pi N},M^2_\Delta,m_\pi^2)}
{2W_{\pi N}}$ and 
$|\vec
p_\pi^{\ \prime\,o-s}|=\frac{\lambda^{1/2}(M_D^2,M^2_\Delta,m_\pi^2)}{2M_D}$.\\

As for the matrix elements of the isoscalar part of the
electromagnetic current associated to the $N^*(1520)$, we make use of
the relations given in Eq.~(\ref{eq:65}) and 
the expression for $\langle n\pi^0|s^\mu_{{\rm em},{\rm IS}}(0)|n\rangle$ 
given in Ref.~\cite{Hernandez:2013jka}. Thus, we obtain
\bea
j^\mu_{\rm em\ IS}|_{DP}&=&iC^{DP}_{\rm IS}g_D\frac1{\sqrt{3}}
\frac{k_\pi^\alpha}{p_D^2-M_D^2+iM_D\Gamma_D}
\bar u(\vec{p}\,' )\gamma_5 \,
P^D_{\beta\alpha}(p_D)\, \Gamma^{D\,\beta\mu}_{V\ {\rm IS}}
\left(p,q \right)
u(\vec{p}\,),\nonumber\\ 
&&\hspace{2cm}p_D=p+q,\ \ C^{DP}_{\rm IS}=\left(\begin{array}{cll}
1&{\rm for}&p\to p\pi^0\\
\sqrt{2}&{\rm for}&p\to n\pi^+\\
-1&{\rm for}&n\to n\pi^0\\
\sqrt{2}&{\rm for}&p\to p\pi^-\\
\end{array}\right),
\eea

\bea
j^\mu_{\rm em\ IS}|_{CDP}&=&-iC^{CDP}_{\rm IS}g_D\frac1{\sqrt{3}}
\frac{k_\pi^\alpha}{p_D^2-M_D^2+iM_D\Gamma_D}
\bar u(\vec{p}\,' )\widehat \Gamma^{D\,\mu\beta}_{V\ {\rm IS}}
\left(p',-q \right)\, 
P_{\beta\alpha}(p_D)\, 
\gamma_5 u(\vec{p}\,),\nonumber\\
&& p_D=p'-q,\ \ C^{DP}_{\rm IS}=\left(\begin{array}{cll}
1&{\rm for}&p\to p\pi^0\\
\sqrt{2}&{\rm for}&p\to n\pi^+\\
-1&{\rm for}&n\to n\pi^0\\
\sqrt{2}&{\rm for}&p\to p\pi^-\\
\end{array}\right),\ \ \widehat \Gamma^{V\,\mu\beta}_{D\,\rm IS}
\left(p',-q \right)=\gamma^0[\Gamma^{V\,\beta\mu}_{D\,\rm IS}
\left(p',-q \right)]^\dag\gamma^0.
\eea
with
\bea
\Gamma^{D\,\beta\mu}_{V\ {\rm IS}}=\left[ 
\frac{\tilde C_3^{V,{\rm IS}}}{M}\left(g^{\beta\mu}
\slashchar{q}-
q^\beta\gamma^\mu\right) + \frac{\tilde C_4^{V,{\rm IS}}}{M^2} \left(g^{\beta\mu}
q\cdot p_D- q^\beta p_D^\mu\right)
+ \frac{\tilde C_5^{V,{\rm IS}}}{M^2} \left(g^{\beta\mu}
q\cdot p- q^\beta p^\mu\right)\right. + \tilde C_6^{V,{\rm IS}} g^{\beta\mu}
\bigg]
\eea
The isoscalar form factors are given in Ref.~\cite{Hernandez:2013jka}. For them
we use the same functional form as for the  $\tilde C^V_j$ while their
values at $q^2=0$  have been taken  from Ref.~\cite{Leitner:2009zz}.

Finally, the differential $\gamma N \to N' \pi$ cross section in the laboratory
(LAB) frame for real
photons is obtained from the amplitudes $j^\mu_{\rm em}$ as
\begin{equation}
\left.\frac{d^2\sigma}{d\cos(\theta_\pi)dE_\pi}\right|_{\rm LAB} = -\frac{\alpha |\vec{k}_\pi|}{16 M |\vec{q}\,|
  E'} \left(\overline{\sum_{\rm spins}} j^\mu_{\rm em}j^*_{\mu\,\rm{em}}\right)
\delta\left(q^0+M-E_\pi-E'\right) \label{eq:cross}
\end{equation}
The energy conservation Dirac delta fixes the pion polar angle in the
LAB frame as
\begin{equation}
\cos(\theta_\pi) = \frac{2M(E_\pi-q^0)+2q^0E_\pi-m_\pi^2}{2q^0|\vec{k}_\pi|}
\end{equation}
In addition, the average and sum over the initial and final nucleon 
spins in Eq.~(\ref{eq:cross}) is readily done thanks to
\begin{equation}
\overline{\sum_{\rm spins}} \bar u(\vec{p}\,') {\cal S^\mu} u(\vec{p}\,) 
\left[\bar u(\vec{p}\,') {\cal S_\mu} u(\vec{p}\,)\right]^* = \frac12
     {\rm Tr}\left((\slashchar{p}^\prime +M){\cal S^\mu}(\slashchar{p}
     +M) \gamma^0{\cal
       S^{\dagger}_\mu}\gamma^0 \right)
\end{equation}
where the spin dependence of the Dirac's spinors is understood and
${\cal S^\mu}$ is a matrix in the Dirac's space for each value of the
Lorentz index $\mu$.
%
%
%
%
%
%
%
%
%
%
%

\end{document}